\documentclass[14pt,twoside,letterpaper]{article} 
\newcommand{\rvec}{\mathrm {\mathbf {r}}} 
\newcommand{\pvec}{\mathrm {\mathbf {p}}}
\newcommand{\nvec}{\mathrm {\mathbf {n}}}
 
\newcommand{\Hvec}{\mathrm {\mathbf {H}}} 
\newcommand{\Vvec}{\mathrm {\mathbf {V}}}

\newcommand{\beq}{\begin{equation}} 
\newcommand{\eeq}{\end{equation}} 
\usepackage{times,fancyhdr}
\usepackage[dvips]{graphicx}
\usepackage{color}
\usepackage{amsfonts}
\usepackage{url}
\usepackage{subfigure}
\usepackage{subfigure}
\usepackage{xcolor}
\usepackage{amsmath}
\usepackage{enumerate}
\usepackage{tabularx}
\usepackage{array}


\makeatletter 
\def\cleardoublepage{\clearpage\if@twoside \ifodd\c@page\else%
    \hbox{}%
    \thispagestyle{empty}%
    \newpage%
    \if@twocolumn\hbox{}\newpage\fi\fi\fi} 
\makeatother

\def\figurename{Figure}
\makeatletter
\renewcommand{\fnum@figure}[1]{\figurename~\thefigure.}
\makeatother

\def\tablename{Table}
\makeatletter
\renewcommand{\fnum@table}[1]{\tablename~\thetable.}
\makeatother

\begin{document}
\title{
{\begin{flushleft}
\vskip 0.45in
{\normalsize\bfseries\textit{Chapter}}
\end{flushleft}
\vskip 0.45in
\bfseries\scshape Information entropy in excited states in confined quantum systems}}

\author{\bfseries\itshape Sangita Majumdar,
\bfseries\itshape Neetik Mukherjee,  
\bfseries\itshape Amlan K. Roy\thanks{E-mail address: akroy@iiserkol.ac.in, akroy6k@gmail.com}\\
Department of Chemical Sciences,\\ 
Indian Institute of Science Education and Research (IISER) Kolkata,\\
Mohanpur-741246, Nadia, India}
\date{} 
\maketitle
\thispagestyle{empty}
\setcounter{page}{1}
\thispagestyle{fancy}
\fancyhead{}
\fancyhead[L]{ pp. {\thepage-\pageref{lastpage-01}}} 
\fancyhead[R]{ISBN 0000000000  \\
\copyright}
\fancyfoot{}
\renewcommand{\headrulewidth}{0pt}

\vspace{2in}


\vspace{.08in} \noindent \textbf{Keywords:} R\'enyi entropy, Shannon entropy, Fisher information, Onicescu energy, Complexities, 
Confined harmonic oscillator, Confined H atom, Confined many-electron atom, Excited state

\clearpage 
\pagestyle{fancy}
\fancyhead{}
\fancyhead[EC]{Sangita Majumdar, Neetik Mukherjee and Amlan K. Roy}
\fancyhead[EL,OR]{\thepage}
\fancyhead[OC]{Information entropy in excited states in confined quantum systems $\cdots$}
\fancyfoot{}
\renewcommand\headrulewidth{0.5pt} 
\begin{abstract}
The present contribution constitutes a brief account of information theoretical analysis in several representative model as well as
real quantum mechanical systems. There has been an overwhelming interest to study such measures in various quantum systems, as 
evidenced by a vast amount of publications in the literature that has taken place in recent years. However, while such works are
numerous in so-called \emph{free} systems, there is a genuine lack of these in their constrained counterparts. With this in 
mind, this chapter will focus on some of the recent exciting progresses that has been witnessed in our laboratory \cite{sen06,roy14mpla,roy14mpla_manning,roy15ijqc, 
roy16ijqc, mukherjee15,mukherjee16,majumdar17,mukherjee18a,mukherjee18b,mukherjee18c,mukherjee18d,majumdar20,mukherjee21,majumdar21a,
majumdar21b}, and elsewhere, with special emphasis on following prototypical systems, namely, (i) double well (DW) potential 
(symmetric and asymmetric) (ii) \emph{free}, as well as a \emph{confined hydrogen atom} (CHA) enclosed in a spherical impenetrable 
cavity (iii) a many-electron atom under similar enclosed environment. 

The information generating functionals, R\'enyi (R) and Tsallis (T) entropies, are 
closely connected to entropic moments, completely characterizing density. Shannon entropy ($S$), Onicescu energy ($E$) are 
two particular cases of $R,T$; former signifying a measure of uncertainty, whereas latter quantifies separation of 
density with respect to equilibrium. Fisher information ($I$) is a gradient functional of density, identifying local fluctuation 
of a space variable. Some lower bounds (which do not depend on quantum numbers) are available for $R, S$; on the other hand,
both upper and lower bounds of $I$ have been achieved, which strictly vary with quantum numbers. Another related concept is 
Kullback-Leibler divergence or \emph{relative entropy}, that delineates how a probability distribution changes from a given reference 
(usually chosen as ground state) distribution. In other words, this quantifies shift of information from one state to another. 
  
A DW oscillator, $v(x)=\alpha x^{m}-\beta x^{n}+\gamma x^{l}$ ($\alpha, \beta, \gamma$ are real numbers, $m,n,l$ are integers 
and $m>n>l$) may be classified into two categories (i) symmetric DW (SDW), where $m,n$ are \emph{even} and $\gamma=0$ (ii) 
asymmetric DW (ADW), where $m,n$ \emph{even}, $\gamma \neq 0$ and $l$ \emph{odd}. They have their relevance in many physical and
chemical phenomena, such hydrogen bonding, umbrella flipping of NH$_3$, anomalous optical lattice vibrations, proton transfer in 
DNA, internal rotation, etc. Through an accurate solution of Schr\"odinger equation by means of a variation-induced exact 
diagonalization procedure involving harmonic oscillator basis, these measures have been calculated. These, along with a phase-space
analysis nicely explains the competing behavior of localization and delocalization. Appearance of \emph{quasi-degeneracy} between 
lower even and adjacent upper odd states can be interpreted using these measures. In ADW a particle in a definite excited 
state can reside either in deeper or shallower well, depending upon the strength of the asymmetric parameter $\gamma$. In this context, 
a pair of rules have been proposed for ADW using energy results and the information analysis. It is found that at certain range of $\gamma$, 
two wells of ADW effectively behave as two different potentials.

Next we would move to the realistic atomic systems. It is interesting to note that the effect of confinement on excited quantum 
states is more dramatic. A detailed investigation of these information quantities along with several \emph{complexity} (in both $r$
and $p$ spaces) would be probed for a caged-in atom. As it appears, some analytical results are possible, which would be considered
accordingly. We will also discuss about the suitability and applicability of a newly designed virial theorem. For many-electron atoms,
the relevant Kohn-Sham equation would be solved by adopting a work-function-based exchange potential, along with a suitable 
correlation energy functional, within the broad domain of \emph{density functional theory}.    

\end{abstract}

\tableofcontents

\section{Introduction}\label{sec:intro}
From the early days of quantum mechanics the problem of a \emph{particle-in-a-box} has played the role of a simple, and 
particularly educative example, showing the difference in energy spectrum between a \emph{free} and a \emph{confined} quantum 
particle. By \emph{free} system, we refer to quantum problems solved in whole space. Analysis of systems in 
some sub-region $\Omega$ of space is important mainly when one tries to model realistic situations in highly inhomogeneous 
media or in intense external fields. But during the last few years, it has become much more than a pedagogical tool and has 
found widespread applications in numerous areas of physics and chemistry. Matter under extreme conditions of pressure 
\cite{jaskolski96, sabin09, sen14, dolmatov04} has been an interesting topic for a long time, for both experimental and 
theoretical researchers \cite{grochala07, michels37, degroot46, leykoo79, leykoo80, marin91, marin92, aquino95, 
aquino09elsevier, katriel12, lesech11, sarsa11, cabrera13}. 
The experimental techniques have given the much needed insight about response of matter under such conditions. 
Because of a wide variety of applications, considerable attention has been given to analyze the effect of very high pressure 
on electronic structure, energy spectrum, chemical reactivity, ionization potential, molecular bond size, polarizabilities of 
atoms and molecules confined by cavities of different geometrical forms and dimension \cite{michels37, jaskolski96, sabin09, 
sen14, connerade00, colin11, montgomery12}, etc. Apart from these, plenty of other applications are also reported, e.g., 
cell model in liquid sate, semiconductor dots, atoms encapsulated within nanocavities, such as fullerene, zeolite molecular 
sieves, solvent environments, porous silicon. Some other systems of interest are confined phonons, polaritons, 
plasmons, confined gas of bosons etc. These models of spatial confinements are also of significance in astrophysics for 
understanding of mass-radius relation of white dwarfs and ionized plasma properties. In order to describe 
states of atoms or molecules in cavities or in semiconductor nanostructures, one could formally perform extensive large-scale 
\emph{first principles} calculations for clusters containing embedded objects. 
In such a case one might get a proper description of the whole system, but still it is difficult to routinely perform such 
elaborate calculations, and may not necessarily lead to simple interpretation of the results gained. While such attempts have 
grown more with time, still it is quite desirable to look for simpler methods to understand confinement effects.

In general, the effect of spatial confinement is of same origin as some other phenomena caused by electric fields in a cavity 
environment. But wherever it is possible, it is usually convenient to consider them separately, distinguishing the barrier-type 
sharp border potentials from other ones. Spatial confinement can thus, in the simplest way, be modeled by putting potential 
barriers at borders of a confining region. In such a scenario, the behavior of wave function at cavity border, and its outside, 
is forced by the presence of the wall. In some occasions, however, potential barriers do not appear directly in equations, 
and we have to impose boundary conditions that reflect the fact that, probability of finding the object outside its region of 
confinement is zero or nearly zero. This means requiring the vanishing (or almost) of the wave function at cavity border, 
which, on the other hand, corresponds to the presence of infinite/finite potential barriers. For a wide range of physical 
situations, one may consider the Schr\"odinger equation for a given subsystem and use an appropriate nontrivial boundary condition
on the boundary $\delta \Omega$ of the concerned region $\Omega$. In order to connect some of the basic concepts, we start by 
pointing out the difference between Dirichlet, Neumann and Robin types. In traditional mathematical physics, one considers the 
following self-adjoint boundary conditions on wave functions:
\begin{eqnarray}
\begin{aligned}
\psi(\rvec)|_{{\rvec} \in \delta \Omega} & = 0 &(\text{Dirichlet boundary condition}),\\
\delta_{n} \psi ({\rvec}) |_{{\rvec} \in \delta \Omega} & = 0 & (\text{Neumann boundary condition}),\\
\delta_{n} \psi ({\rvec}) - P({\rvec}) \psi({\rvec}) |_{{\rvec} \in \delta \Omega} & = 0 & (\text{General boundary condition}),
\end{aligned}
\end{eqnarray}
where $\delta_{n}$ is the normal derivative, i.e., $\delta_{n} \psi = (\nvec, \nabla \psi)$, for a unit vector $\nvec$  
externally normal to the surface $\delta \omega$ at a given point, and $\nabla \psi$ is the gradient of wave function $\psi$, 
while $P({\rvec})$ denotes some real-valued function. The Neumann boundary condition implies $P \equiv 0$, while for Dirichlet 
boundaries one may suppose formally $|P| \rightarrow \infty$. 

Probably one of the first papers that analyzed an atomic system in a bounded region, was due to Wigner and Seitz \cite{wigner33, 
wigner34} on the theory of periodic structures, published in 1933-1934. The Schr\"odinger equation for an atom in a lattice was 
studied with Neumann boundary conditions. In this work $\Omega$ corresponded to a sphere of radius $R$ and the nucleus was placed 
at its center. This helped reduce the problem to that of radial function only. Later, Michels \emph{et al.,} \cite{michels37} 
considered a hydrogen atom confined within a sphere built with rigid walls, such that the electron density, $\rho(\rvec)$, vanishes 
at the boundary of confining radius. This publication is often attributed as first example of how Dirichlet boundary condition
can be applied to study real physical problems. Note that the approach is still in use, especially in astrophysics. Prior to this
work, however, in 1911, Weyl solved some vibrational problems, which may be interpreted as describing the structure of highly 
excited part of the spectrum of a particle in a bounded region $\Omega$ with Dirichlet boundary conditions. 

Confinement model is very useful to examine effects on atoms trapped in a microscopic cavity. In order to investigate 
pressure and polarizabilty as a function of compression, a model consisting of a hydrogen atom at the centre of an impenetrable box, 
was proposed \cite{michels37}, which has become very popular in dealing with confined quantum systems with boxes of different size 
and geometrical forms. Similarly, many-electron atoms subjected to extreme pressure can also be simulated by placing them
inside an impenetrable cavity of adjustable radius. Here, the infinite potential is induced by neighboring particles of negative 
charge. In literature, this kind of trapping is known as hard confinement. The electrostatic Hamiltonian is modified by adding 
a confining potential in term of radius $r_c$. However, this  only includes effects produced by repulsive forces. To account for 
the existence of attractive forces between particles, such as van der Waals forces, it was proposed that the potential surface 
be finite. Confined atoms by penetrable confining potentials were also scrutinized by modifying adequately the extra potential 
added to the Hamiltonian. It may be emphasized that the incorporation of a particular boundary condition poses significant difficulties 
in calculation of the corresponding energy spectra. Hence the success of a given method in ground state may not necessarily be carried 
forward in excited states. Recent progresses in understanding of caged atoms taking into account various confining 
environments and their modeling, changes in their properties, as well as the methods of analysis and solutions 
along with their accuracy, are illustrated in the next section.

An immense amount of theoretical works have been published over several decades, covering a broad variety of confined systems. 
It has been analyzed by putting the atom inside boxes of different geometrical forms (with varying size within hard boundary); the 
most preferred choice 
is a spherical box of penetrable or impenetrable walls. A confined H atom (CHA) within an impenetrable \cite{sobrino17, goldman92, 
aquino95, garza98, laughlin02, burrows06, aquino07, baye08, ciftci09, roy15ijqc, roy16ijqc} (as well as penetrable) cavity was 
investigated quite vigorously, as recognized to be an original treasure-trove of numerous attractive properties, both from physical 
and mathematical point of view. 
Finding approximate analytical as well as numerical solutions, and comparing their accuracy to exact solution has been an active 
field of research. A broad range of theoretical methods varying in difficulty, sophistication and accuracy is available; a 
selected set includes perturbation theory, Pa\'de approximation, WKB method, Hypervirial theorem, power-series solution, 
super-symmetric quantum mechanics, Lie algebra, Lagrange-mesh method, asymptotic iteration method, generalized pseudospectral 
method \cite{sobrino17, goldman92, aquino95, garza98, laughlin02, burrows06, aquino07, baye08, ciftci09, roy15ijqc, 
roy16ijqc} etc., and the references therein. Exact solution \cite{laughlin02} of CHA has been put forth
in terms of Kummer M-function (confluent hypergeometric). While CHA energy spectrum shows a monotonic, unlimited increase of its energies 
as the boundary approaches nucleus and volume of confinement is reduced, 
the same in case of open conoidal boundaries (sphere, circular cone, paraboloid, prolate spheroid, hyperboloid) is characterized 
by a monotonic increase of energy levels only up to zero energy in the corresponding limit situations, with consequent infinite 
degeneracy. The exact solution of Schr\"odinger equation for hydrogen atom in spherical, spheroconal, parabolic and prolate 
spheroidal coordinate has also been reported by several researchers. In contrast to the free hydrogen atom (FHA), in CHA due to 
symmetry breaking, different energy eigenvalues, eigenfunctions and reduced degeneracies is observed. Study of CHA with soft 
spherical boxes can also be found in literature where the compression regimes are considered with Neumann boundary conditions. 
They offer many unique phenomena, especially relating to \emph{simultaneous, incidental and inter-dimensional degeneracy} 
\cite{sen05,mukherjee21}. Effect of compression on ground and excited energy levels, as well as other properties, like hyper-fine 
splitting constant, dipole shielding factor, nuclear magnetic screening constant, pressure, static, and dynamic polarizability, were 
probed. 

We now extend our discussion to many-electron atoms confined in various environments. The analysis of such systems are complicated 
due to the electron-electron repulsion term. Unlike hydrogen, in such systems, the electron 
correlation, which is an essential component of electronic structure, can be studied. As confinement by hard impenetrable boxes 
overestimates the response of physical observables, penetrable walls are more convenient to mimic the experimental counterpart.
Confined atoms centered within a sphere enveloped by \emph{hard walls}--a widely used model to imitate enclosed environment, can be 
considered as a starting point to the understanding of confinement inside \emph{penetrable walls} \cite{montgomery12, sabin09, 
colin11, laughlin09, gorecki88, rodriguez15, sarsa15}. Helium atom being the prototypical many-electron atom and one of the most 
abundant elements found to exist as a significant component of giant planets and stars, is a most relevant one to approach the 
electronic structure and spectroscopic properties, when subjected to high pressure. Calculating electronic properties of these 
systems, in which electronic clouds are forced to remain spatially restricted, represents a more demanding task. The investigations 
on He confinement was carried out fifteen years after 
the study of hydrogen atom by Seldam \emph{et al.} \cite{tenseldam52} where they performed a variational calculation with Hylleraas 
wave function, multiplied by a cut-off function, for ground state. Thereafter, several other analyses have been extended to determine 
the electronic structure of ground as well as few low lying excited states, which can be broadly categorized as: i) wave-function based 
and ii) density functional theory (DFT) techniques. The Hartree-Fock (HF) and Kohn-Sham (KS) models are the most popular and promising  
routes to obtain wave function and electron density, although variational approaches based on the generalized Hylleraas basis set 
(where the wave function contains a cut-off factor to ensure Dirichlet boundary condition) has also been employed towards their 
understanding \cite{aquino06, wilson10, flores10, flores08, sakiroglu09}. There are basically two ways to obtain the wave function 
or electron density under these boundary conditions: using a cut-off function multiplied with basis functions in restricted HF 
framework to impose such a confinement \cite{lesech11, sarkar09, vanfaassen09} or imposing the behavior on wave function within 
the numerical scheme when solving SCF method for the radial equation \cite{connerade00, garza98}. Another alternative variational 
method is the so-called optimized effective potential approximation, which has been generalized to work with multi-configuration 
wave function. This was applied to compute electronic structure of both ground and excited states in constrained atoms \cite{sarsa14, 
sarsa16, galvez17}. Several other noteworthy techniques that were engaged towards these are: Rayleigh-Ritz \cite{gimarc67, marin92}, 
linear variational \cite{rivelino01}, HF \cite{ludena78, connerade00}, configuration interaction \cite{ludena79}, quantum Monte 
Carlo \cite{joslin92}, etc. Here it is worth mentioning that, although there exist several wave-function based methods for 
these kind of systems, density-based attempts are rather very scarce \cite{garza98, aquino06}. Apart from these, considering 
the nucleus to be fixed at the 
centre of spherical box, Whitkop \cite{whitkop99} achieved a novel contribution upon studying with nucleus placed off centre of the 
impenetrable spherical box.

In the last few decades, the concept of quantum information theory has emerged as a subject of topical interest due to its extensive  
applications \cite{poland2000,singer04,anton07} in understanding various phenomena in physics and chemistry, relating to diverse 
topics such as thermodynamics, quantum mechanics, spectroscopy etc. In recent years, it has also been used prodigiously in the 
study of quantum entanglement and quantum steering problems. In literature, the information-theoretic measures like Shannon 
entropy ($S$) \cite{birula75,shannon51}, R\'enyi entropy ($R^{\alpha}$) \cite{birula06}, Fisher information ($I$), Onicescu energy 
($E$) and complexities ($C$) are invoked to explore multitude of phenomena such as, diffusion of atomic orbitals, spread of electron 
density, periodic properties, correlation energy and so forth. Arguably, entropic uncertainty relations, constructed from these 
fundamental information theoretical tools, are the most effective measures of uncertainty \cite{birula75,birula06, duan2000, 
simon2000}, as they do not make any reference to some specific points of the respective Hilbert space. In a quantum system, $S$ 
and $I$ quantify the information content in a complimentary way. Former is a global measure of spread of density which refers to the 
expectation value of logarithmic probability density function. On the other hand, $I$ is a gradient functional of density and it 
signifies the oscillatory nature and sharpness of density in position space ($r$). It is well known that, $R^{\alpha}, T^{\alpha}$, 
the so-called information generating functionals, are closely connected to entropic moments (discussed later), and completely 
characterize density $\rho(\rvec)$. They actually quantify the spatial delocalization of single-particle density of a system in 
various complimentary ways. Moreover, these are closely related to energetic and experimentally measurable quantities 
\cite{gonzalez03prl,sen12}. It is interesting to note that, $S$, $E$ (disequilibrium) are two particular cases of 
$R^{\alpha}, T^{\alpha}$ \cite{sen12,birula06}. Former measures total extent of density whereas $E$ quantifies separation of 
density with respect to equilibrium. However, $(-)$ve value in $R$ and $S$ indicates extreme localization, whereas, $E$ is always 
($+$)ve. Likewise, changing the numerical values of $R^{\alpha},~S$ from $(-)$ve to ($+$)ve only interprets enhancement of 
de-localization. Another associated concept is \emph{complexity}. A system has finite complexity when it is either in a state with 
less than some maximal order, or not at a state of equilibrium. In a nutshell, it becomes \emph{zero} at two limiting cases, 
\emph{viz.,} when a system is (i) completely ordered (maximum distance from equilibrium) or (ii) at equilibrium (maximum disorder). 
Complexity has its contemporary interest in chaotic systems, spatial patterns, language, multi-electronic systems, molecular or DNA 
analysis, social science, astrophysics and cosmology etc \cite{sen12}. 

A particle in an SDW potential is a prototype of a system for which perhaps, a full quantum mechanical 
explanation of confinement is more effective than that of traditional uncertainty product relation. This has 
relevance in unraveling a wide variety of physical, chemical phenomena of a model two-state system \cite{verguilla93}, wherein one 
may identify two wells as two different states of a quantum particle. Some important applications worth recording are, hydrogen 
bonding, umbrella flipping of ammonia molecule, anomalous optical lattice vibrations 
in HgTe, proton transfer in DNA to model brain  micro-tubules, internal rotation, physics of solid-state devices, solar cells and 
electron tunneling microscopes, etc. Existence of an interplay between localization and delocalization effects is another interesting 
feature of this system. These 
competing effects lead to a number of quasi-degenerate pair states. An increase in positive term (strength) reduces spacing 
between classical turning points but at the same time, reduces barrier area as well as barrier height. Conversely, negative term 
increases spacing between classical turning points but also increases barrier area and barrier height. It has been pointed out that 
the characteristic feature of quasi-degeneracy present in a DW potential can not be examined from 
commonly used uncertainty relation. 

Introduction of an asymmetric term in the DW potential makes it more anomalous and interesting. Because 
it gives rise to two asymmetric wells. From a classical mechanical point of view, a particle will always reside in 
deeper well, but in quantum mechanics the situation is not that straightforward. 
Some typical applications of asymmetric DW potential are: electron in a double quantum dot, Bose-Einstein 
condensation in trapped potentials, quantum super-conducting circuit based on Josephson junction, 
quantum computing devices, etc. The contrasting effects present in such systems, especially in excited states, can be 
beautifully analyzed in terms of information entropic measures.

Next, in this paragraph, we shift towards the information in atoms. In recent years, appreciable attention has been paid 
to investigate various information measures in central potentials. At first we restrict ourselves to a few references pertaining to 
H atom--both FHA and CHA. Some of these for FHA are: $I_{r}, I_{p}, I$ in 3D \cite{romera05}, in D-dimension \cite{romera06,dehesa06, 
dehesa07}; upper bounds of $S, R$ \cite{sanchez11}, $S$ in 3D \cite{toranzo16phy}, in D-dimension \cite{yanez94}; $R, T$ in 3D 
\cite{toranzo16phy}, in D-dimension \cite{toranzo16epl}. Relativistic effects on the information measures of FHA are also examined 
\cite{katriel10}. A lucid review on information theory of D-dimensional FHA is provided in \cite{dehesa10}. 
 The exact mathematical form of $I$ for an arbitrary state of FHA was given in both r and p 
space \cite{romera05} in terms of four expectation values $\langle r^2 \rangle, \langle p^{-2} \rangle, \langle p^2 \rangle, 
\langle p^{-2} \rangle$, and eventually can be expressed in terms of related quantum numbers. Likewise, an exact analytical formula 
for $S$ in ground state of a D-dimensional FHA was derived long times ago \cite{yanez94} in both r and p space. Later, similar 
analytical expressions of $S$ for circular or node-less states of a D-dimensional FHA was offered in 2010 \cite{dehesa10} in both 
spaces. However, such a closed form expression of $S$ is as yet lacking for a general state. Very recently, in \cite{jiao17}, accurate 
radial Shannon entropies in $r, p$ spaces, with $n=10$ are computed numerically. Moreover, a generalized form of angular Shannon 
entropy was also derived. Recently, $R$ and $T$ for Rydberg hydrogenic states were reported within a strong Laguerre asymptotic 
approximation. \cite{toranzo16phy, toranzo16epl}. But their exact closed-shell forms are as yet unknown; moreover these were produced 
so far, only in $r$ space, and mostly for $l=0$ states. For FHA, all these can be calculated from exact analytical wave functions in 
$r, p$ spaces. Recently we have found that, for node-less states, expressions of $S, R, T, E$ are accessible in closed form in a FHA, 
as the required radial polynomial reduces to unity. But for all other $n, l$ they need to be computed numerically, as the presence of 
nodes in such wave functions leads to difficult polynomials. 

However, analogous studies as above, in CHA are quite scanty and are of recent origin; e.g., $S$ \cite{sen05,jiao17,mukherjee18c}, bounds of 
$I, S, R, T$ \cite{patil07}, as well as $I, S$ in case of soft spherically CHA. \cite{aquino13}.  
Apart from the work \cite{jiao17} for $S$, it was analyzed in the context of soft and hard confinement in lowest state 
\cite{sen05, aquino13}. The variation of $S$ in $s, p, d$ orbitals $(n \leq 7)$ was followed \cite{jiao17} with $r_c$. A detailed 
systematic variation of these measures in CHA, with respect to $r_c$ (as well as FHA) has been presented only 
lately \cite{mukherjee18c, mukherjee18d}; for non-zero angular momentum they are separately published in \cite{mukherjee18e}. 
hydrogen$-$like ions in conjugate $r$ and $p$ spaces has been done in our lab. \cite{mukherjee18d}. Besides these independent 
information quantities, several well-known measures similar to LMC and Fisher-Shannon complexity have been pursued for CHA in both 
conjugate spaces \cite{majumdar17}. 

When one analyzes these aforementioned quantities for many-electron systems, electron correlation comes into play. By treating the 
electron probability distribution as a continuous function, these quantities are applied in the study of different model potentials 
mimicking diatomic molecular potentials \cite{majumdar19} and numerous atomic and molecular systems \cite{guevara03, 
guevara05, romera08, nagy09a, antolin09, antolin09b, zhou16, toranzo16epl, toranzo16phy, dehesa17}. Amongst these, $S$ is the 
hallmark of information theory \cite{nagy13,ou19}. From the standpoint of not being able to obtain any analytic forms, significant 
amount of numerical study has been published for innumerable many-body systems. For instance, an entropy maximization technique was 
applied to explain the Compton profiles of atoms \cite{sears81}. Later, Gadre \emph{et al.} \cite{gadre84} considered $S$ within a
Thomas-Fermi model. They also speculated that Shannon entropy sum, $(S_{t})$ can be used as an indicator of quality of N-electron 
wave functions \cite{gadre85}. This was extended for similar works \cite{dehesa89} in arbitrary D-dimensional many-particle system. 
A validation came from an investigation \cite{tripathi92} of electronic correlation in Be-isoelectronic series. A numerical HF 
calculation \cite{sen12} for free neutral atoms suggested that $S_t$ could play the role of an indicator of shell structure.
Beyond HF-level calculations were reported as well \cite{ho94, tripathi96}. $S$ has various other applications as 
well in the study of free atoms, namely, descriptor of several properties of a finite coulomb system in both ground and excited states 
\cite{nagy13}, measure of aromaticity \cite{noorizadeh10}, manifests in avoided crossing \cite{gonzalez03prl}, quantifies amount of 
electronic interactions present in a system \cite{sagar02}, measure of correlation \cite{guevara03} and so on. $I$  has been demonstrated 
to be a very useful tool in analyzing atoms \cite{nalewajski03, nagy03, romera05, nagy06a, nagy06, romera04, romera02, liu07, nagy07, 
nagy08, szabo08}. Moreover, $S$, $I$ have been depicted as an indicator of quality of an approximate wave function \cite{nagy03}. 
Though a large-scale of study for these quantities are available in literature for free atoms, amount of similar study for 
confined systems is very limited. Sen \cite{sen05} has examined $S$ for H-and He-like systems with impenetrable 
boundary. The author tried to locate if there exists any optimum point in $S_t$ as a function of confinement radius. Ou 
\emph{et al.} \cite{ou17} pursued similar analysis for He confined with penetrable potential in its ground and excited 
states by employing highly correlated and extensive Hylleraas-type wave function. HF study of electron-delocalization 
in other many-electron atoms confined by penetrable walls in terms of $S$ has also been executed \cite{rodriguez18}.
The present authors have investigated these information theoretic measures for atoms and ions under high pressures in 
both ground and excited states employing a work function based strategy within the framework of DFT \cite{majumdar21,majumdar21a,
majumdar21b}. 

Besides the aforementioned information tools, complexity measures also have the ability to estimate a variety of physical or 
biological systems for organization. Statistical measure of complexity were presented for atoms using HF densities in both 
$r$ and $p-$ spaces where analysis reveals interesting correlations with quantities such as ionization potential, 
static dipole polarizability, shell structure of electron density etc. \cite{sen14}. Apart from one-electron densities, two-electron 
atomic densities (pair density) were also studied in terms of complexity measures (shape complexity) and its corresponding information 
planes \cite{lopez19}. In addition to atomic systems, it has applications in diverse fields like detection of periodic, quasi-periodic, 
linear stochastic, chaotic dynamics etc.

The present contribution gives an outline of the recent progresses that have taken place in the information theoretical analysis of 
quantities, like $S, R, I, E, C,$ in the ground and excited states of several quantum confined systems, such
as a particle confined in SDW and ADW potential, CHA. Furthermore, we have also dealt with a KS model, which has been 
applied successfully for calculation of energy 
spectrum of many-electron atoms under similar kind of confinement. In the 1D DW potential, we introduce a variation-induced exact 
diagonalization method for the accurate solution of relevant eigenvalue equation developed in our laboratory. In the 3D counterpart 
for confined atomic systems, we employ the very accurate generalized pseudospectral (GPS) method. These solutions are used throughout 
the chapter for the calculation of desired quantities presented. Detailed results are compiled in tabular and graphical form. Their changes 
are carefully monitored as one goes from unbounded to bounded system. Additionally we also offer a brief account of the relative 
information studies that has been undertaken by us recently. Available literature results have been consulted as and when possible. 
A few concluding remarks are made at the end. 

\section{Theoretical Aspects}
This section provides the required details for theoretical solution of relevant eigenvalue equations as well as calculation of 
information-related quantities, as probed in this chapter. To simplify the discussion, three categories are made. At first, in order 
to achieve accurate eigenvalues and eigenfunctions in 1D DW potentials, a variation induced exact diagonalization method, involving 
manifold energy minimization scheme is adopted in both $x$ and $p$ spaces. In the second class, we deal with CHA, whole exact wave 
functions are available in the form of Kummer confluent hypergeometric function; respective energies are obtained via accurate  
GPS method. Finally, in the context of multi-electron atoms, we present a work function-based strategy; here also we engage the GPS 
method for solution of KS equation. 

\subsection{Variation induced exact diagonalization}
It is a simple basis-set method, and in principle, exact energies can be achieved provided the basis is complete. For convenience, 
the Hermite basis is adopted. Then the Hamiltonian can be represented in terms of raising and lowering operator of a conventional 
quantum harmonic oscillator (QHO), characterized by a potential of the form $V(x)=4 \sigma^2 x^2$. Thus, a QHO number operator basis 
with a single non-linear parameter $\sigma$ ($\sigma$ is related to the force constant $k$ as $\sigma^{2}=\frac{1}{8}k$) was utilized. 
In all the calculations that follow, dimension of the Hamiltonian matrix is set at $N=100$ to ensure convergence of eigenvalues. 
Employing such a near-complete basis, matrix elements of Hamiltonian are defined as, 
\begin{equation}
h_{mn}=\langle m| \hat{H} | n \rangle. 
\end{equation}   
Diagonalization of the matrix $h$ provides the energy eigenvalues and corresponding eigenvectors, which is  easily performed by 
MATHEMATICA package.

In this work, instead of minimizing the individual eigenvalue, a manifold-energy minimization method \cite{hendekovic83,pathak94} was 
invoked. In this scheme, the trace of the matrix with respect to $\sigma$ is minimized, which remains invariant under diagonalization 
of the matrix. Hence, one needs to diagonalize the matrix at that particular $\sigma$ value for which trace of the matrix is minimum. 
This leads to all desired eigenvalues in a single diagonalization step. 

The eigenvectors in position space then can be written in following form, 
\begin{equation}
\psi_n(x)=\sum_{m=0}b_{m}^n\left(\frac{2\sigma}{\pi}\right)^
{\frac{1}{4}}\frac{1}{\sqrt{2^{m}m!}} \ H_{m}(\sqrt{2\sigma}x) \ e^{-{\sigma}x^{2}}
\end{equation}
while in momentum space, they are manifested as,
\begin{equation}
\psi_n(p)=\sum_{m=0}c_{m}^n\left(\frac{1}{2\sigma\pi}\right)^{\frac{1}{4}}\frac{1}{\sqrt{2^{m}m!}}
\ H_{m} \left( \frac{p}{\sqrt{2\sigma}} \right) \ e^{-\frac{p^{2}}{4\sigma}}.
\end{equation}

\subsubsection{Symmetric double-well (SDW) potential}
For our purposes, the SDW potential is conveniently written as, 
\begin{equation} \label{sdw}
V(x) = \alpha x^{4}-\beta x^{2} + \frac{\beta^{2}}{4 \alpha}.
\end{equation}
It is symmetric around $x = 0$ and two minima are located at $x_{0} = \pm \sqrt{\frac{\beta}{ 2 \alpha}}$, while the maximum value 
of potential is $\frac{\beta^2}{4\alpha}$ \cite{jelic12}. An increase in quartic parameter, $\alpha$ shortens 
separation between classical turning points and reduces barrier strength. Thus, on one side, it promotes localization and on 
other hand, minimizes confinement on the particle within a well. A rise in $\beta$ causes duelling effects on 
particle. It leads to an enhancement in separation of classical turning points implying added delocalization of particle, 
whereas at the same time, barrier area and barrier height increase, promoting localization into one of the wells.
 
The Hamiltonian in position space is expressed as:
\begin{equation}
\hat{H_{x}}=-\frac{d^{2}}{dx^{2}}+\alpha x^{4}-\beta x^{2} + \frac{\beta^{2}}{4 \alpha},
\end{equation}
whereas in momentum space, this reads as below,  
\begin{equation}
\hat{H_{p}}=p^{2}+\alpha\frac{d^{4}}{dp^{4}}+\beta\frac{d^{2}}{dp^{2}}+ \frac{\beta^{2}}{4 \alpha}.
\end{equation}
For sake of convenience, we choose $m=\frac{1}{2}$ and $\hbar=1$.

In position space, the non-zero matrix elements are expressed as \cite{mukherjee15},
\begin{eqnarray}
h_{lm} & = & \frac{\alpha}{16\sigma^{2}}\sqrt{l(l-1)(l-2)(l-3)}, \hspace{2.05in} \mathrm{if} \  l-m=4  \nonumber \\
       & = & \left[\frac{\alpha}{8\sigma^{2}}(2l-1)-\frac{(\beta+4\sigma^{2})}{4\sigma}\right]\sqrt{l(l-1)}, \hspace{1.43in} \mathrm{if} \  l-m=2 \nonumber \\
       & = & \left[2\sigma(2l+1)-(\beta+4\sigma^{2})\frac{(2l+1)}{4\sigma}+\frac{3\alpha}{16\sigma^{2}}(2l^{2}+2l+1)\right], \hspace{0.22in} \mathrm{if} \  l-m=0 \nonumber \\
       & = & \left[\frac{\alpha}{8\sigma^{2}}(2l-1)-\frac{(\beta+4\sigma^{2})}{4\sigma}\right]\sqrt{(l+1)(l+2)}, \hspace{1.03in} \mathrm{if} \  l-m=-2 \nonumber \\
       & = & \frac{\alpha}{16\sigma^{2}}\sqrt{(l+1)(l+2)(l+3)(l+4)}. \hspace{1.65in} \mathrm{if} \  l-m=-4  \nonumber \\
\end{eqnarray}
Conversely, in $p$ space the non-zero terms are expressed in the following forms \cite{mukherjee15}, 
\begin{eqnarray}
g_{lm} & = &  h_{lm}, \hspace{2.2in}  \mathrm{if} \  (l-m)=4,-4  \nonumber \\
       & = & -h_{lm},  \hspace{2.07in}  \mathrm{if} \  (l-m)=2,-2 \nonumber  \\
       & = &  h_{lm},   \hspace{2.2in}  \mathrm{if} \  (l-m)=4,-4   
\end{eqnarray} 
The trace of $h/g$ matrix in either spaces at a particular $\sigma$ is same, given by, 
\begin{equation}
Tr[h/g]=\sum_{m}h_{mm} = \sum_{m}\left[\frac{3\alpha}{16\sigma^{2}}(2m^{2}+2m+1)-
\frac{(\beta + 4\sigma^2)(2m+1)}{4\sigma}+2\sigma(2m+1)\right].
\end{equation}  
Now, minimization of the trace leads to a cubic equation in $\sigma$,
\begin{equation}\label{gen_eq}
8{\sigma^3}+2\sigma\beta-\alpha \ \frac{(2{N^2}+4N+3)}{(N+1)}=0.
\end{equation}
Use of parity, transforms Eq.~(\ref{gen_eq}), for even $m$ as,
\begin{equation}\label{min_even}
8{\sigma^3}+2\sigma\beta-\alpha(2N+1)=0,
\end{equation}
while for odd $m$ this offers, 
\begin{equation}\label{min_odd}
8{\sigma^3}+2\sigma\beta-\alpha(2N+3)=0. 
\end{equation}
It can be easily demonstrated that $\sigma$ will have a single real root in Eqs.~(\ref{gen_eq}), (\ref{min_even}) and (\ref{min_odd}), 
for which the trace will be minimum for respective cases. Now, one can diagonalize the matrix employing the value of non-linear 
parameter $\sigma$.

\subsubsection{Asymmetric double well (ADW) potential}
\begin{figure}                        
\begin{minipage}[c]{1.0\textwidth}\centering
\includegraphics[scale=0.87]{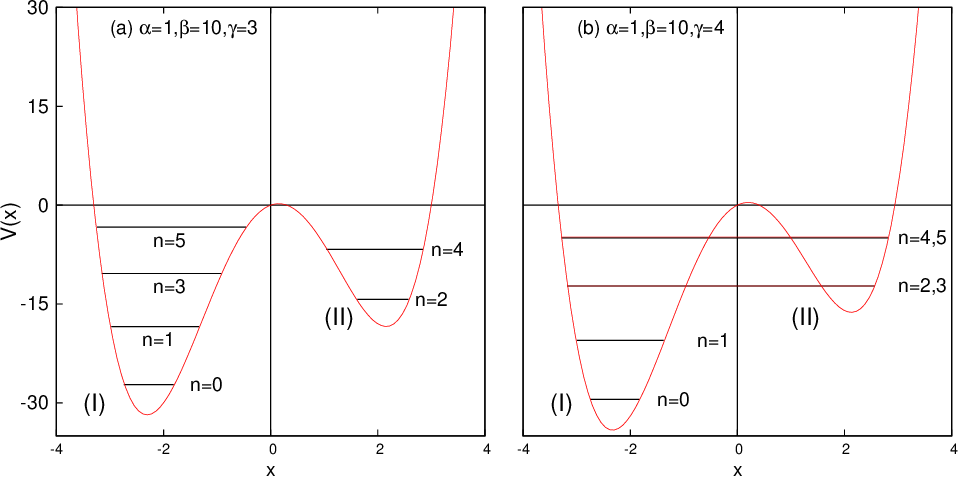}
\end{minipage}%
\caption{Schematic representation of an ADW potential, in Eq.~(\ref{pot}), plotted at two different parameter sets: (a) 
$\alpha \! = \! 1, \beta \! = \! 10, \gamma \! = \! 3$ (b) $\alpha \! = \! 1, \beta \! = \! 10, \gamma \! = \! 4$. (I), 
(II) signify larger, smaller wells respectively. See text for details.}
\label{asdw} 
\end{figure}

Without any loss of generality, the potential is simply written as \cite{mukherjee16}, 
\begin{eqnarray} \label{pot}
V(x) = \alpha x^{4}-\beta x^{2} + \gamma x + V_0,\\
V_{1}(x) = \alpha x^{4}-\beta x^{2} - \gamma x + V_0.
\end{eqnarray}
Here $V_0$ signifies the values of global minimum of the potential. It has been added to make total energy positive. Clearly, $V$ 
and $V_{1}$ constitute a mirror-image pair; substitution of $x$ by $-x$ transforms one to other. These two potentials are 
\emph{isospectral} in nature. Interestingly, their wave functions are also mirror images to each other. Thus, it is adequate to 
study the behavior of any one of them; other one automatically follows. This prompts us to choose $V(x)$ as our model ADW potential. 
Incorporation of a linear term in a  SDW brings asymmetry in the system. Note that, $\alpha$ and $\beta$ exert analogous effect to 
what we have observed in SDW potential. Introduction of asymmetric term shifts the maximum of potential from zero to either right or 
left sides depending on whether $\gamma$ is positive or negative. Apparently, there exists a deeper (I) and a shallow (II) well. 
Figure~\ref{asdw} illustrates two typical scenario at two different $\gamma$ values. It also portrays that maximum of $V$ shifts 
towards right as $\gamma$ increases. 

The Hamiltonian in position space (for simplicity, we choose $2m \! = \! 1, \hbar=1$) is:
\begin{equation}
\hat{H_{x}}=-\frac{d^{2}}{dx^{2}}+\alpha x^{4}-\beta x^{2} + \gamma x,
\end{equation}
whereas in momentum space, this reads, 
\begin{equation}
\hat{H_{p}}=p^{2}+\alpha\frac{d^{4}}{dp^{4}}+\beta\frac{d^{2}}{dp^{2}}-i\gamma\frac{d}{dp}.
\end{equation}

In position space, the non-zero matrix elements can be written as follows \cite{mukherjee16}, 
\begin{eqnarray}
h_{lm} & = & \frac{\alpha}{16\sigma^2}\sqrt{l(l-1)(l-2)(l-3)}, \hspace{2.05in} \mathrm{if} \  l-m=4  \\
        & = & \frac{1}{8\sigma^2}[\alpha(2l-1)-2(\beta+4\sigma^2) \sigma]\sqrt{l(l-1)}, \hspace{1.3in}  
\mathrm{if} \ l-m=2  \nonumber \\
        & =& \gamma \sqrt{\frac{l}{4\sigma}}, \hspace{3.65in} \mathrm{if} \ l-m=1  \nonumber  \\
        &= & \frac{3\alpha}{16\sigma^2}(2l^2+2l+1)-(\beta+4\sigma^2) \frac{(2l+1)}{4\sigma}+2 \sigma (2l+1),  
\hspace{0.35in} \ \mathrm{if} \ l-m=0   \nonumber \\
        & = & \gamma \sqrt{\frac{(l+1)}{4\sigma}}, \hspace{3.35in}  \mathrm{if} \ l-m=-1 \nonumber \\
        & = & \frac{1}{8\sigma^2}[\alpha(2l+3)-2(\beta+4\sigma^2) \sigma]\sqrt{(l+1)(l+2)},  \hspace{0.95in}  
\mathrm{if} \ l-m=-2 \nonumber \\
        & = & \frac{\alpha}{16\sigma^2}\sqrt{(l+1)(l+2)(l+3)(l+4)}.  \hspace{1.78in} \mathrm{if} \ l-m=-4  \nonumber 
\end{eqnarray}
On the other hand, in momentum space, the non-zero elements are obtained as \cite{mukherjee16}, 
\begin{eqnarray}
g_{lm} & = &  h_{lm}, \hspace{2.2in}  \mathrm{if} \  (l-m)=4,-4  \nonumber \\ 
       & = & -h_{lm}, \hspace{2.05in} \mathrm{if} \  (l-m)=2,-2 \nonumber \\ 
       & = &-ih_{lm}, \hspace{2.05in}    \mathrm{if} \  (l-m)=1 \nonumber    \\
       & = & ih_{lm}. \hspace{2.12in} \mathrm{if} \  (l-m)=-1  \nonumber \\
       & = &  h_{lm}, \hspace{2.2in}  \mathrm{if} \  (l-m)=0 
\end{eqnarray}
Diagonalization of the \emph{symmetric} matrix, {\bf h/g} was performed efficiently by MATHEMATICA, leading to accurate energy eigenvalues and 
corresponding eigenvectors. We adopt a Manifold-Energy minimization approach \cite{hendekovic83,pathak94}, where instead of minimizing a 
particular energy state, one minimizes trace of the matrix, which is given below, 
\begin{equation} \label{trace}
Tr[h]=\sum_{l}h_{ll} = \sum_{l}\left[\frac{3\alpha}{16\sigma^{2}}(2l^{2}+2l+1)-
\frac{(\beta + 4\sigma^2)(2l+1)}{4\sigma}+2\sigma(2l+1)\right], 
\end{equation}
with respect to $\sigma$. This leads to a cubic equation in $\sigma$ having only one real root. Finally the process is accomplished by 
minimizing the desired matrix in Eq.~(\ref{trace}), for above value of $\sigma$.

\subsection{Confined atomic system}
In this subsection, at first we discuss CHA and then a many-electron atom. The desired confinement is achieved by shifting the radial 
boundary from \emph{infinity} to a \emph{finite} range.
\subsubsection{Position-space wave function}
\paragraph{CHA:}
The exact radial wave function for a CHA can be expressed \cite{burrows06} as, 
\begin{eqnarray} \label{wfh}
&  &  \psi_{n, l}(r)  =  N_{n, l}\left(2r\sqrt{-2\mathcal{E}_{n,l}}\right)^{l} \times  \\ \nonumber 
&  & _{1}F_{1} \left[\left(l+1-\frac{1}{\sqrt{-2\mathcal{E}_{n,l}}}\right),(2l+2),2r\sqrt{-2\mathcal{E}_{n,l}}\right]      
e^{-r\sqrt{-2\mathcal{E}_{n,l}}},
\end{eqnarray}
where $N_{n, l}$ indicates normalization constant and $\mathcal{E}_{n,l}$ is the energy of a given state characterized by $n,l$ quantum 
numbers, whereas $_1F_1\left[a,b,r\right]$ illustrates confluent hypergeometric function. In case of FHA ($r_c \rightarrow \infty$), the 
first-order hypergeometric function modifies to associated Laguerre polynomial with $\mathcal{E}_{n,l}=-\frac{Z^2}{2n^2}$ ($Z$ denotes 
atomic number). So the radial function simplifies to commonly used form, as given below, 
\begin{equation}
\psi_{n,l}(r)= \frac{2}{n^2}\left[\frac{(n-l-1)!}{(n+l)!}\right]^{\frac{1}{2}}\left[\frac{2Z}{n}r\right]^{l} 
e^{-\frac{Z}{n}r} \ L_{(n-l-1)}^{(2l+1)} \left(\frac{2Z}{n}r\right).  
\end{equation}
Thus allowed energies at a specific $r_c$ can be obtained by finding the zeros of $_{1}F_{1}$, 
\begin{equation}
_{1}F_{1}\left[\left(l+1-\frac{1}{\sqrt{-2\mathcal{E}_{n,l}}}\right),(2l+2),2r_{c}
\sqrt{-2{\mathcal{E}_{n,l}}}\right]=0. 
\end{equation}
At a certain $l$, the first root corresponds to energy of the lowest-$n$ state $(n_{lowest}=l+1)$ with gradual roots identifying higher 
excited states. It is worthwhile noting that, in order to construct the exact wave function of CHA for a specific state, one requires to 
supply energy eigenvalue of that state. In our calculation, $\mathcal{E}_{n,l}$ of CHA, computed by means of the GPS method. The most 
important feature is that through an optimal, non-uniform spatial grid, it offers a symmetric eigenvalue problem, which can be easily 
solved by standard diagonalization routines, offering accurate eigenvalues and eigenvectors. Over the years, it has been applied to a
variety of model and real systems, in both free and confined quantum systems, viz., spiked harmonic oscillators \cite{roy04pla, roy08theochem}, 
power-law and logarithmic \cite{roy04jpg}, H\'ulthen and Yukawa \cite{roy05pramana}, rational \cite{roy08ijqc}, Hellmann \cite{roy08jmc}, 
exponential-screened Coulomb \cite{roy13}; various molecular potentials for ro-vibrational states such as Morse \cite{roy13rinp}, 
hyperbolic \cite{roy14fbsys}, shifted Deng-Fan \cite{roy14ijqc}, Tietz-Hua \cite{roy14jmc}, Manning-Rosen \cite{roy14mpla_manning}, 
as well as other singular \cite{roy05ijqc} potentials. Very successful applications have also been made in the context of many-electron 
systems within the broad domain of DFT \cite{roy02b, roy02qfd, roy02, roy04jpbhollow, roy05jpbhollow, roy07}. In past several years, 
this has also produced excellent quality results in various radial confinement studies in 3DQHO \cite{sen06, roy14mpla, roy15ijqc, roy16ijqc}. 

At this moment, it is necessary to point out that, there exists an incidental degeneracy in CHA under following condition:
{\bf{For all $n \ge (l+2)$, a $(n,l)$ state is degenerate with $((n+1),(n+2))$ state at $r_{c}=(l+1)(l+2)$.}} For example, 
at $r_{c}=2$, energies of $1s$ and $3d$ states of CHA are degenerate; so are $2p$ and $4f$ states at $r_{c}=6$  

\paragraph{Many-electron atom:}
The starting point is the single-particle time-independent Kohn-Sham (TIKS) equation, which can be conveniently written as (atomic units
employed, unless mentioned otherwise), 
\begin{equation}
    \Hvec_{0}(\rvec)\psi_i(\rvec)=\epsilon_{i}(\rvec)\psi_{i}(\rvec) ,
\end{equation}
where $\Hvec_0$ is the unperturbed-KS Hamiltonian, given by
\begin{eqnarray}
    \Hvec_0(\rvec) = -\frac{1}{2}\nabla^{2}+\Vvec_{eff}(\rvec) \nonumber \\
    \Vvec_{eff}(\rvec) = v_{ne}(\rvec) +\int \frac{\rho(\rvec^{\prime})}{|\rvec - \rvec^{\prime}|}
 \mathrm{d}\rvec^{\prime}+\frac{\partial E_{xc}[\rho(\rvec)]}{\partial \rho(\rvec)} .  
\end{eqnarray}
Though DFT has achieved impressive success in explaining electronic structure and properties of many-electron system in their ground state, 
calculation of excited-state energies and densities has remained a bottleneck. In this work, we employ a work-function method \cite{sahni92}, 
according to which, exchange energy can be interpreted as interaction energy between an electron at $\rvec$ and its Fermi-Coulomb hole 
charge density $\rho_{x}(\rvec,\rvec^{\prime})$, $\rvec$ at $\rvec^{\prime}$,
\begin{equation}
 E_{x}[\rho(\rvec)] = \frac{1}{2}\int\int\frac{\rho(\rvec)\rho_{x}(\rvec,\rvec^{\prime})}{|\rvec - \rvec^{\prime}|} \ 
\mathrm{d}\rvec\mathrm{d}\rvec^{\prime}
\end{equation}
The exchange potential can then be defined as work done in bringing an electron to the point $\rvec$, against an electric field
produced by its Fermi-Coulomb hole density.
\begin{equation}
\epsilon_{x}(\rvec) = \int\frac{\rho(\rvec)\rho_{x}(\rvec,\rvec^{\prime})(\rvec - \rvec^{\prime})}
{|\rvec - \rvec^{\prime}|^{3}}
\end{equation}
The above expression gives the electric field due to Fermi-Coulomb hole charge density and from that
the exchange potential can be obtained as,
\begin{equation}
v_{x} (\rvec) = -\int_{\infty}^{r} \epsilon_{x}(\rvec) \mathrm{d}l
\end{equation}
While the exchange potential $\mathit{v}_x(\rvec)$ can be accurately calculated, the correlation potential $\mathit{v}_c(\rvec)$ is unknown 
as yet, and must be approximated. We have taken into consideration two correlation energy functionals such as local Wigner \cite{brual78} 
and non-local Lee-Yang-Parr (LYP) \cite{lee88} functional. With this choice, this equation is solved numerically with GPS method. Now, to study 
the effect of hard confinement, we have imposed Dirichlet's boundary condition. The solution of TIKS equation obtained 
gives orthonormal atomic orbitals, from which the one-particle density can be found as, 
\begin{equation}
 \rho(\rvec) = \sum_{\i=1}^{N} |\psi_i(\rvec)|^2.  
\end{equation}
Now information measures can be calculated in conjugate spaces, by standard procedure.  
 
\subsubsection{Momentum-space wave function}
The $p$-space wave function ($\pvec = \{ p, \Omega \}$) of a particle in a central potential is achieved from respective 
Fourier transform of its $r$-space counterpart \cite{mukherjee18c}, and is given below,
\begin{equation} \label{eq:psi_p}
\begin{aligned}
\psi_{n,l}(p) & = & \frac{1}{(2\pi)^{\frac{3}{2}}} \  \int_0^\infty \int_0^\pi \int_0^{2\pi} \psi_{n,l}(r) \ \Theta(\theta) 
 \Phi(\phi) \ e^{ipr \cos \theta}  r^2 \sin \theta \ \mathrm{d}r \mathrm{d} \theta \mathrm{d} \phi,  \\
      & = & \frac{1}{2\pi} \sqrt{\frac{2l+1}{2}} \int_0^\infty \int_0^\pi \psi_{n,l} (r) \  P_{l}^{0}(\cos \theta) \ 
e^{ipr \cos \theta} \ r^2 \sin \theta  \ \mathrm{d}r \mathrm{d} \theta.  
\end{aligned}
\end{equation}
Note that $\psi(p)$ needs to be normalized. Integrating over $\theta$ and $\phi$ variables, leads to, 
\begin{equation}
\psi_{n,l}(p)=(-i)^{l} \int_0^\infty \  \frac{\psi_{n,l}(r)}{p} \ f(r,p) \ \mathrm{d}r.    
\end{equation}
Depending on $l$, this can be rewritten in following form ($m'$ starts with 0),  
\begin{equation}
\begin{aligned}
f(r,p) & = & \sum_{k=2m^{\prime}+1}^{m^{\prime}<\frac{l}{2}} a_{k} \ \frac{\cos pr}{p^{k}r^{k-1}} +  
  \sum_{j=2m^{\prime}}^{m^{\prime}=\frac{l}{2}} b_{j} \ \frac{\sin pr}{p^{j}r^{j-1}}, \ \ \ \ \mathrm{for} \ 
  \mathrm{even} \ l,   \\
f(r,p) & = & \sum_{k=2m^{\prime}}^{m^{\prime}=\frac{l-1}{2}} a_{k} \ \frac{\cos pr}{p^{k}r^{k-1}} +  
\sum_{j=2m^{\prime}+1}^{m^{\prime}=\frac{l-1}{2}} b_{j} \ \frac{\sin pr}{p^{j}r^{j-1}}, \ \ \ \ \mathrm{for} \ \mathrm{odd} \ l.
\end{aligned} 
\end{equation}
The co-coefficients $a_{k}$, $b_{j}$ of even- and odd-$l$ states can be found in Tables~1 and 2 of \cite{mukherjee18c}. 

\section{Formulation of Information-theoretical quantities}
In this section we shall briefly discuss the various information-theoretic quantities along with their mathematical forms. This will 
provide the context where these quantities are defined and the relations between them.

\subsection{Shannon entropy ($S$)}
Information is mobile and can be carried over from one place to another. The basic idea of information theory is that, the more one knows 
about an event, the less new information one is apt to get about it. If an event is very probable, there is much less uncertainty and 
thus it offers little new information. In summary, the information content is an increasing function of inverse of the probability of 
an event, and proportional to the uncertainty existing before its occurrence \cite{ birula75,cover06,nielsen10}.  
\begin{equation}
\begin{aligned}
\mathrm{Information~received~by~occurrence~of~an~event~} = \mathrm{Amount~of~uncertainty~} \\
\mathrm{prevailed~before~its~occurrence}
\end{aligned}
\end{equation} 
Shannon in 1948 connected the measure of information content with a discrete probability distribution ($p_{1}, p_{2},p_{3},...,p_{n}$) 
consisting of $n$ different events, by giving a measure quantifying the uncertainty in results, in the given form \cite{shannon51},
\begin{equation}
S=-k\sum{p_{i}~\mathrm{ln}~p_{i}}, 
\end{equation}
where $k$ is a positive constant depending on the choice of unit. This definition can be explained by choosing two limiting cases: (i) at 
first, when any of the $p_{i}=1$ and others are \emph{zero}, then for this certain event $S=0$, which is minimum (ii) in case of 
equi-probability, where all $p_{i}=\frac{1}{n}$, and the uncertainty of the outcome is maximum, then $S$ is also maximum ($S=k\ln n$) 
\cite{cover06}. In essence, it can be said that, for a given distribution, lesser the probability of occurring an event, higher will be the 
uncertainty associated with it. Hence, after occurrence of that event, more information will come out. Potentially, $S$ is claimed as the 
best measure of information \cite{birula75}.

The concept of $S$ has been extensively used in wave mechanics to describe various phenomena, such as Colin conjecture \cite{ramirez97, 
delle15}, atomic avoided crossing \cite{ghiringhelli10, ghiringhelli10a}, orbital-free DFT \cite{nagy14, alipour15}, electron correlation 
\cite{delle09, mohajeri09, grassi11, gallegos16, alipour18}, configuration-interaction, entanglement in artificial atoms 
\cite{barghathi18}, aromaticity \cite{noorizadeh10} in many-electron systems, etc. Additionally, the Shannon entropy sum $(S_t=S_{\rvec} + 
S_{\pvec})$ which obeys well known BBM inequality and contains net information, can be proved to be a stronger version of the traditional 
Heisenberg uncertainty principle in quantum mechanics \cite{birula75}. The entropic uncertainty relation has the mathematical form 
\cite{birula75},
\begin{equation} \label{eq:20} 
S_{\rvec}+S_{\pvec} \ge D(1+\mathrm{ln}~\pi), 
\end{equation}  
where $D$ refers to dimensionality of the system. Here, $S_{\rvec}$, $S_{\pvec}$ have following forms,
\begin{equation}
\begin{aligned} 
S_{\rvec}  & =  -\int_{{\mathcal{R}}^3} \rho(\rvec) \ \ln [\rho(\rvec)] \ \mathrm{d} \rvec   = 
2\pi \left(S_{r}+S_{(\theta,\phi)}\right), \\   
S_{\pvec}  & =  -\int_{{\mathcal{R}}^3} \Pi(\pvec) \ \ln [\Pi(\pvec)] \ \mathrm{d} \pvec   = 
2\pi \left(S_{p}+S_{(\theta, \phi)}\right), \\ 
S_{r} & =  -\int_0^\infty \rho(r) \ \ln [\rho(r)] r^2 \mathrm{d}r, \ \ \ \ \ \ \ \ \rho(r) = |\psi_{n,l}(r)|^{2}, \\
S_{p} & =  -\int_{0}^\infty \Pi(p) \ln [\Pi(p)]  \ p^2 \mathrm{d}p, \ \ \ \ \ \ \ \ \Pi(p) = |\psi_{n,l}(p)|^{2}, \\
S_{(\theta, \phi)}  & =   -\int_0^\pi \chi(\theta) \ \ln [\chi(\theta)] \sin \theta \mathrm{d} \theta, \ \ \ \ \ \
\chi(\theta)   =  |\Theta(\theta)|^2.  \\  
\end{aligned} 
\end{equation}

\subsection{R\'enyi entropy ($R$)}
R\'enyi entropy is a one-parameter extension of family of information measures having several similar properties to $S$. 
It was introduced in 1961 as a generalized version of $S$ which for a given probability distribution $P=(p_{1},p_{2},....,p_{n})$ 
\cite{renyi61} can be defined as,
\begin{equation}
R_{\alpha}=\frac{1}{1-\alpha} \ln \left(\sum_{k} p_{k}^{\alpha}\right).
\end{equation} 
It is to be noted that, $R_{\alpha}$ is considered as an alternative measure of information because 
(i) $R_{\alpha}$ is the exponential mean of information entropy whereas $S$ provides the arithmetic mean of it, and (ii) 
at $\alpha \rightarrow \infty$, $R_{\alpha}$ reduces to $S$ \cite{birula75}. It is an information generating functional of 
$\alpha$-order entropic moments, which can completely characterize density $\rho(\rvec)$. In case of continuous density distribution, 
it may be expressible in terms of expectation values of density, in following standard form \cite{renyi61,renyi70},
\begin{equation}
R^{\alpha}[\rho (\rvec)]  =  \frac{1}{(1-\alpha)} \ \mathrm{ln} \ \langle \rho(\rvec)^{(\alpha-1)}\rangle,  \ \ \ \ 0 < \alpha 
< \infty, \ \ \alpha \neq 1. 
\end{equation}
Entropic uncertainty relations usually carry more information over the commonly used quantum mechanical uncertainty relations such as, 
$\Delta x \Delta p \ge \hbar$. While 
the latter signifies the area of a phase space ($\delta \rvec \delta \pvec$) obtained by resolution of measuring instruments, the former 
does not provide us with any such information. It suggests that with an increase of localization of the particle in phase space, the 
sum of uncertainties in position and momentum space escalates. The quantum mechanical uncertainty relation containing the phase space is 
of the form,
\begin{equation} \label{eq:21}
R_{\rvec}^{\alpha}+R_{\pvec}^{\alpha} \ge -\frac{1}{2}\left(\frac{\ln~\alpha}{1-\alpha}+\frac{\ln~\beta}{1-\beta}\right) - 
\ln~\frac{\delta \rvec \delta \pvec}{\pi \hbar};  \ \ \ \ \ \left(\frac{1}{\alpha}+\frac{1}{\beta}\right)=2.
\end{equation} 
In the limit, when $\alpha \rightarrow 1$ and $\beta \rightarrow 1$, this uncertainty relation reduces to Eq.~(\ref{eq:20}). 
However, the above relation presents a better insight about uncertainty as it contains all-order entropic moments \cite{birula06}. But 
still further improvement is required, which remains an open challenging problem. Interestingly, Eq.~(\ref{eq:21}) becomes sharper and 
sharper when the relative size of phase-space area $\left(\frac{\delta \rvec \delta \pvec}{\pi \hbar}\right)$ defined by the experimental 
resolution reduces. In fact, it \emph{is the case}, as one enters more and more in to the quantum territory.        

In quantum mechanics, $R^{\alpha}$ has been successfully employed to investigate and speculate many quantum phenomena 
such as, entanglement, communication protocol, correlation de-coherence, measurement, localization properties of Rydberg 
states, molecular reactivity, multi-fractal thermodynamics, production of multi-particle in high-energy collision, disordered 
systems, spin system, quantum-classical correspondence, localization in phase space \cite{varga03,renner05,levay05,verstraete06,
bialas06,salcedo09,nagy09,mcminis13,liu15,kim18}, etc. 

R{\'e}nyi entropies of order $\lambda (\neq 1)$ ($\lambda$ is either $\alpha$ or $\beta$) are obtained by taking logarithm of 
$\lambda$-order entropic moment \cite{renyi70,birula06}. In spherical polar coordinate these can be expressed in following 
simplified form, by means of some straightforward algebra \cite{mukherjee18c}, 
\begin{equation}
\begin{aligned} 
R_{\rvec}^{\lambda}  =  \frac{1}{1-\lambda} \ln \left(\int_{{\mathcal{R}}^3} \rho^{\lambda}(\rvec)\mathrm{d} \rvec \right)  = &
\frac{1}{(1-\lambda)} \ln \left(2\pi\int_0^\infty [\rho(r)]^{\lambda} r^2 \mathrm{d}r \int_0^\pi [\chi(\theta)]^{\lambda} \sin 
\theta \mathrm{d}\theta \right) \\ 
 = & \frac{1}{(1-\lambda)}\left( \ln 2\pi + \ln [\omega^{\lambda}_r] + \ln [\omega^{\lambda}_{(\theta, \phi)}] \right),  \\
R_{\pvec}^{\lambda}  =  \frac{1}{1-\lambda} \ln \left[\int_{{\mathcal{R}}^3} \Pi^{\lambda}(\pvec)\mathrm{d} \pvec \right]  = &
\frac{1}{(1-\lambda)} \ln \left(2\pi \int_{0}^\infty [\Pi(p)]^{\lambda} p^2 \mathrm{d}p \int_0^\pi [\chi(\theta)]^{\lambda} \sin 
\theta \mathrm{d}\theta \right) \\
 = & \frac{1}{(1-\lambda)}\left( \ln 2\pi + \ln [\omega^{\lambda}_p] + \ln [\omega^{\lambda}_{(\theta, \phi)}] \right).
\end{aligned} 
\end{equation}          
Here $\omega^{\lambda}_{\tau}$s are entropic moments in $\tau$ ($r$ or $p$ or $\theta$) space with order $\lambda$, having forms,
\begin{equation}
\omega^{\lambda}_r= \int_0^\infty [\rho(r)]^{\lambda} r^2 \mathrm{d}r, \ \ \  
\omega^{\lambda}_p= \int_{0}^\infty [\Pi(p)]^{\lambda} p^2 \mathrm{d}p, \ \ \  
\omega^{\lambda}_{(\theta, \phi)}= \int_0^\pi [\chi(\theta)]^{\lambda} \sin \theta \mathrm{d}\theta. 
\end{equation}          

\begin{figure}                          
\begin{minipage}[c]{1.0\textwidth}\centering
\includegraphics[scale=0.78]{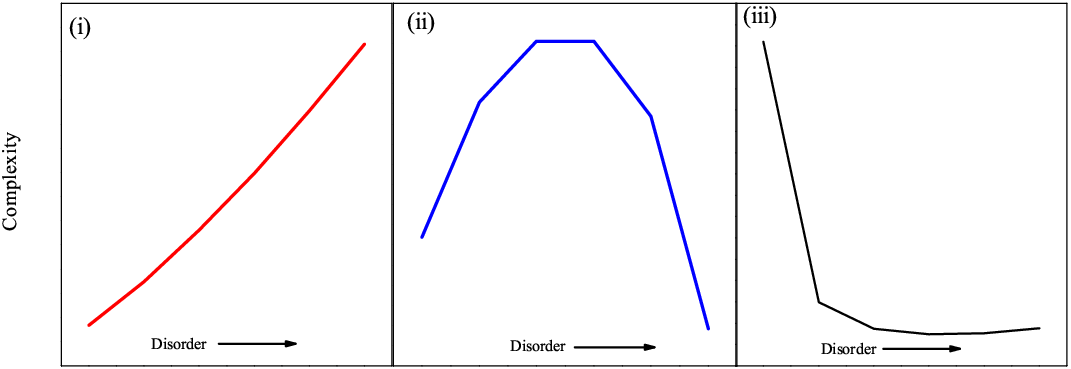}
\end{minipage}%
\caption{Three different categories of Complexities, as functions of disorder.}
\label{comp}
\end{figure} 

\subsection{Fisher information ($I$)}
The idea of entropy can satisfactorily explain the degree of disorder of a given phenomenon. 
Besides, however, it is required to find out a suitable measure of disorder whose variation derives the event. 
The concept of entropy is incapable of providing this. But, having the ability to approximate a parameter, $I$ 
can serve as a good candidate for this purpose. Thus, it becomes a cornerstone of the statistical field of study, known as
parameter estimation \cite{frieden04}. If $e^{2}$ be the mean-square error in an estimation of $\hat{\theta}$, 
then $I$ measures the expected error in accordance with the following relation,
\begin{equation} \label{eq:22}
\begin{aligned}
e^{2}I \ge 1; \ \ \ \
I \ge \frac{1}{e^{2}}.
\end{aligned}
\end{equation}
Equation~(\ref{eq:22}) suggests that, $I$ is always greater than the reciprocal of $e^{2}$; only in case of \emph{Gaussian} distribution it 
becomes equal to inverse of $e^{2}$. The general form of $I$ is,
\begin{equation}
I=\int \frac{|\nabla \rho(\tau)|^{2}}{\rho(\tau)} \ \mathrm{d}\tau, 
\end{equation}     
which is a gradient functional of density quantifying the local density fluctuation in a given space. While for a sharp 
distribution $I$ is higher, it is lower for flat distribution. Thus, it is evident that with a rise in uncertainty, $I$ decreases. It 
resembles the famous Weizs\"acker kinetic energy functional, $T_{\omega}[\rho]$, frequently used in DFT \cite{sen11,nagy14}. 

For central potential $I_{\rvec}$ and $I_{\pvec}$, the net Fisher information, in $\rvec$ and $\pvec$ spaces respectively, 
are expressed as \cite{romera05},
\begin{equation}
\begin{aligned} \label{fish_cp}
I_{\rvec}  =  \int_{{\mathcal{R}}^3} \left[\frac{|\nabla\rho(\rvec)|^2}{\rho(\rvec)}\right] \mathrm{d}\rvec  =  
4\langle p^2\rangle - 2(2l+1)|m|\langle r^{-2}\rangle \\ 
I_{\pvec} =  \int_{{\mathcal{R}}^3} \left[\frac{|\nabla\Pi(\pvec)|^2}{\Pi(\pvec)}\right] \mathrm{d} \pvec  = 
4\langle r^2\rangle - 2(2l+1)|m|\langle p^{-2}\rangle, \\
\langle r^2\rangle=\int_{0}^{r_c} \psi_{n, l}^{*}(r) r^{4} \ \psi_{n, l}(r) \mathrm{d}r, \  \  \ \ \
\langle p^2\rangle=\int_{0}^{r_c} \psi_{n, l}^{*}(r) [-\nabla^{2}\psi_{n, l}(r)] r^2 \mathrm{d}r  \\
\left\langle \frac{1}{r^2}\right\rangle=\int_{0}^{r_c} \psi_{n, l}^{*}(r) \psi_{n, l}(r) \mathrm{d}r, \  \  \ \ \
\left\langle \frac{1}{p^2}\right\rangle=\int_{0}^{r_c} \psi_{n, l}^{*}(p) \psi_{n, l}(p) \mathrm{d}p. 
\end{aligned}
\end{equation}
The above equations can be further modified in following equivalent forms \cite{mukherjee18d, mukherjee18},
\begin{equation}
\begin{aligned}
I_{\rvec} & = 8\mathcal{E}_{n,l}-8\langle v(r)\rangle-2(2l+1)|m|\langle r^{-2}\rangle \\
I_{\pvec} & = 8\mathcal{E}_{n,l}-8\langle T\rangle-2(2l+1)|m|\langle p^{-2}\rangle \\
          & = \frac{8}{\omega^{2}}\langle v(r)\rangle-2(2l+1)|m|\langle p^{-2}\rangle,
\end{aligned}
\end{equation}
where $v(p)$ is the $p$-space counterpart of $v(r)$. 

When $m=0$, $I_{\rvec}$ and $I_{\pvec}$ in Eq.~(44) can be further simplified into,  
\begin{equation}
\begin{aligned} 
I_{\rvec}  =  4\langle p^2\rangle, \hspace{3mm} \ \ \ \  I_{\pvec} =4\langle r^2\rangle.
\end{aligned}
\end{equation}
It is seen that, at fixed $n, l$, both $I_{\rvec}, I_{\pvec}$ provide maximum values when $m=0$, and both of them 
decline with rise in $m$. Hence the upper bound for $I_t$ can be obtained as,
\begin{equation}
 I_{\rvec} I_{\pvec} \ (=I_t) \leq 16 \langle r^2\rangle \langle p^2\rangle .
\end{equation}
Further adjustment using Eq.~(45) leads to following uncertainty relations \cite{romera05}, 
\begin{equation}
\frac{81}{\langle r^2\rangle \langle p^2\rangle} \leq I_{\rvec} I_{\pvec} \leq 16 \langle r^2\rangle \langle p^2\rangle.  
\end{equation}
Therefore, in a central potential, $I$-based uncertainty product is state dependent. It changes with variation in 
$n$, $l$ and is bounded by both upper as well as lower limits.
 
\subsection{Onicescu energy ($E$)}
In 1966, Onicescu introduced the concept of information energy $(E)$, in an endeavour to provide a finer measure of dispersion distribution 
than that of $S$. $E$, for a discrete probability distribution can be defined as, 
\begin{equation}
E=\sum_{i} p_{i}^{2}.
\end{equation} 
which can be extended for a continuous distribution as (expectation value of probability density),
\begin{equation}
E=\int \rho^{2}(\tau) \ \mathrm{d}\tau.
\end{equation}
Therefore, in this occasion, disequilibrium and information energy have same form. This quantity shows a reverse trend to $S$; greater the 
information energy, more concentrated is the probability distribution and the information content reduces. Similar to the previous 
measures, it is also utilized in orbital-free DFT \cite{alipour15}, testing normality \cite{noughabi15}, electron correlation 
\cite{gallegos16}, Colin conjecture \cite{ramirez97,delle15}, configuration interaction \cite{alcoba16} etc.  

By definition, $E$ refers to the 2nd-order entropic moment \cite{sen11}; for central potential it assumes the form below ($E_{t}$ is the 
Onicescu energy product),
\begin{equation}
\begin{aligned}
E_{r} & = \int_0^\infty [\rho(r)]^{2} r^2 \mathrm{d}r, \ \   
E_{p}  = \int_{0}^\infty [\Pi(p)]^{2} p^2 \mathrm{d}p, \\
E_{\theta, \phi} & = \int_0^\pi [\chi(\theta)]^{2} \sin \theta \mathrm{d}\theta.  \ \
E_{t} = E_{r}E_{p}E_{\theta, \phi}^{2}.
\end{aligned} 
\end{equation}
Uncertainty product for such measures are studied in \cite{zozor07}.

\subsection{Complexities}
Simplified systems or idealizations are always a starting point to solve scientific problems. From the most basic grounds, based on our 
common knowledge, a system is said to be ``complex" when it deviates from a pattern regarded as simple. An atom is itself a complex system 
and confinement introduces more complexity to it. Complexity arises in a system due to breakdown of certain symmetry rules. Quantitative 
study of this measure gives an idea of organization in a system and can be considered as a general descriptor of structure and correlation. 
Due to its dependence on the nature of description or on the scale of observation, no univocal characterization of this quantity is 
available. Some notable ones to be mentioned from the definitions available in literature are: Shiner, Davidson and Landsberg (SDL) 
\cite{landsberg98, shiner99}, Fisher-Shannon $(C_{IS})$ \cite{romera04, sen07, angulo08}, Cram\'er-Rao \cite{angulo08, antolin09}, 
generalized R\'enyi-like \cite{calbet01, martin06, romera08} complexity, etc. They can be divided into three broad categories, depending 
on their behavior: 
(i) advances monotonically with disorder (ii) reaches its minimal value for both completely ordered and disordered systems, and a maximum 
at some intermediate level (iii) increases with order. It has finite value in a state lying between two limiting cases of complete order 
(maximum distance from equilibrium) and maximum disorder (at equilibrium). Figure~\ref{comp} pictorially depicts above three different kinds 
of complexities as functions of disorder.  

The ``statistical measure of complexity ($C_{LMC}$)" is one such measure which is based on the statistical description of a system. In 
product form this can be written as, $C_{LMC} = H. D$ where $H$ is nothing but the product of information content (such as $S$, $R$ etc.) 
and $D$ is the concentration of spatial distribution. This quantity was later criticized \cite{catalan02} and modified \cite{sanchez05} to 
the form of $C_{LMC} = D. e^{S}$, to satisfy few conditions such as reaching minimal values for both extremely ordered and disordered limits, 
invariance under scaling, translation and replication. Interestingly, amongst the above mentioned complexities, $C_{IS}$ corresponds to a 
measure which probes a system in terms of complementary global and local factors, and also satisfies certain desirable properties \cite{sen11}, 
like invariance under translations and re-scaling transformations, invariance under replication, near-continuity, etc. This has remarkable 
applications in the study of atomic shell structure, ionization processes \cite{sen07,angulo08,angulo08a}, as well as in molecular properties 
like energy, ionization potential, hardness, dipole moment in the localization-delocalization plane showing chemically significant pattern 
\cite{esquivel10}, molecular reactivity studies \cite{welearegay14}. Some elementary chemical reactions such as hydrogenic-abstraction 
reaction \cite{esquivel11}, identity $SN^{2}$ exchange reaction \cite{molina12}, and also concurrent phenomena occurring at the transition 
region \cite{esquivel12} of these reactions have been investigated through composite information-theoretic measures in conjugate spaces.

Without any loss of generality, let us define complexity in following general form $C = Ae^{b.B}$. The order ($A$) and disorder 
parameters ($B$) may include ($E, I$) and ($R, S$) respectively. With this in mind, we are interested in the following four
quantities, 
\begin{equation} \label{complexity}
\begin{aligned}
C_{ER} & = E e^{bR}, \ \ \ \ \ \ \ \ C_{IR} = Ie^{bR}, \ \ \ \ \ \ \ C_{ES} & = E e^{bS}, \ \ \ \ \ \ \ \ C_{IS} = Ie^{bS}. 
\end{aligned} 
\end{equation}

\begin{figure}             
\centering
\begin{minipage}[c]{1.0\textwidth}\centering
\includegraphics[scale=0.90]{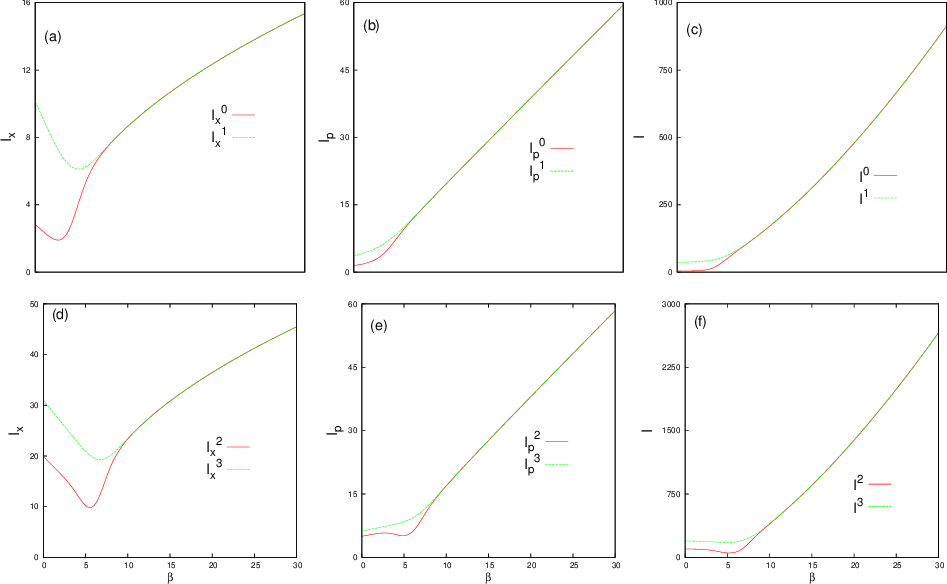}
\end{minipage}%
\caption[optional]{Convergence of $I_x, I_p, I$ against $\beta$ at $\alpha=1$ for SDW potential, Eq.~(\ref{sdw}). Left, 
middle, right columns correspond to $I_x$, $I_p$, $I$ respectively. Top panels (a)-(c) refer to ground and 
first excited states, while bottom panels (d)-(f) refer to third, fourth excited states \cite{mukherjee15}.}
\label{sdw_fish}
\end{figure}

\subsection{Relative information}
Kullback-Leibler divergence or relative entropy is a descriptor of deviation of a given probability distribution from a reference 
reference one \cite{kullback51,kullback78}. In quantum mechanics, this characterizes a measure of distinguishability between two 
states quantifying the change of information from one state to another \cite{frieden04}. 
Relative $R$ and $S$ were studied for various atoms using H atom ground-state as reference \cite{sagar08}. 
Their direct relation with atomic radii and quantum capacitance has been reported \cite{nagy09a,sagar08}. Relative Fisher information 
(IR) is another such interesting measure \cite{villani00}. It has witnessed considerable applications in physics and chemistry, 
such as, in the calculation of phase-space gradient of dissipated work and information \cite{yamano13}, deriving Jensen 
divergence \cite{sanchez12}, relation with score function \cite{toscani17}, in the context of probability current 
\cite{yamano13a}, in thermodynamics \cite{frieden99}, etc. It has also been used in formulating atomic densities 
\cite{antolin09a} and deriving density functionals, under local-density and generalized-gradient approximations \cite{levamaki17}. 
Further, IR along with Hellmann-Feynman and virial theorem was used to develop a Legendre transform structure related to 
SE \cite{flego11}. On the basis of estimation theory, it was designed in a self-consistent manner \cite{frieden10}. 
In quantum chemistry perspective, the above two theorems and entropy maximization principle were used in its formulation \cite{flego11a,
venkatesan14,venkatesan15}. Very recently, IR for some exactly solvable potentials including 1D and 3D QHO in both position and 
momentum spaces was evaluated analytically using a ground state of definite symmetry (for example, $l=0$ for $s$ orbital, $l=1$ 
for $p$ orbital, and so on) as reference \cite{mukherjee18b}. For H-atom using $1s$ orbital as basis, IR in position 
space was numerically estimated \cite{yamano18}. 

For two normalized probability densities $\rho_{n,l,m}(\tau), \rho_{n_{1},l_{1},m_{1}}(\tau)$, IR can be written as, 
\begin{equation}
\mathrm{IR} \ [\rho_{n,l,m}(\tau)|\rho_{n_{1},l_{1},m_{1}}(\tau)]=\int_{{\mathcal{R}}^3}\rho_{n,l,m}(\tau)\left|\nabla \ 
\mathrm{ln} \left\{ \frac{\rho_{n,l,m}(\tau)}{\rho_{n_{1},l_{1},m_{1}} (\tau)}\right \} \right|^{2}  \mathrm{d}\tau.
\end{equation}
Here $n,l,m$ and $n_{1},l_{1},m_{1}$ are the identifiers of target and reference states respectively, while $\tau$ is a 
generalized variable. Without any loss of generality, in case of two central potentials, these probability densities
 $\rho_{n,l,m}(\tau)$, $\rho_{n_{1},l_{1},m_{1}}$ can be conveniently expressed as,
\begin{equation}
\begin{aligned}
\rho_{n,l,m}(\tau) & = R_{n,l}^{2}(s) \Theta_{l,m}^{2}(\theta);  \    \ \ \ \ \ \ \  
\rho_{n_{1},l_{1},m_{1}}(\tau) = R_{n_{1},l_{1}}^{2}(s) \Theta_{l_{1},m_{1}}^{2}(\theta). 
\end{aligned}
\end{equation}
In the above equation, $R_{n,l}(s), R_{n_{1},l_{1}}(s)$ signify radial parts, where $``s"$ implies either $r$ or 
$p$ variable and $\Theta_{n,l}(\theta), \Theta_{n_{1},l_{1}} (\theta)$ represent angular contributions of two wave functions. 
Thus Eq.~(53) may be recast in the form,
\begin{multline}
\mathrm{IR} \ [\rho_{n,l,m}(\tau)|\rho_{n_{1},l_{1},m_{1}}(\tau)]= \mathrm{IR} \ [\rho_{n,l}(s)|\rho_{n_{1},l_{1}}(s)]+
\left\langle \frac{1}{s^{2}} \right\rangle \mathrm{IR} \ [\Theta_{l,m}^{2}(\theta)|\Theta_{l_{1},m_{1}}^{2}(\theta)]+\\
2\int_{0}^{\infty}s~R_{n,l}^{2}(s)\left[\frac{d}{ds}\mathrm{\ln}\left\{ \frac{R^{2}_{n,l}(s)}{R^{2}_{n_{1},l_{1}}(s)}\right\}
\right]\mathrm{d}s
\int_{0}^{\pi}\Theta_{l,m}^{2}(\theta)\left[\frac{d}{d\theta}\mathrm{ln}\left\{ \frac{\Theta_{l,m}^{2}(\theta)}
{\Theta_{l_{1},m_{1}}^{2}(\theta)}\right\} \right] \sin\theta \mathrm{d}\theta
\end{multline}
where the following quantities have been defined, 
\begin{equation} \label{ir}
\begin{aligned}
\mathrm{IR} \ [\rho_{n,l}(s)|\rho_{n_{1},l_{1}}(s)]& =\int_{0}^{\infty}s^{2}~\rho_{n,l}(s)\left|\frac{d}{ds}\mathrm{ln}
\left\{ \frac{\rho_{n,l}(s)}{\rho_{n_{1},l_{1}}(s)}\right\} \right|^{2}\mathrm{d}s \\ 
\mathrm{IR} \ [\Theta_{l,m}^{2}(\theta)|\Theta_{l_{1},m_{1}}^{2}(\theta)]&=\int_{0}^{\pi}\Theta_{l,m}^{2}(\theta)\left|
\frac{d}{d\theta}\mathrm{ln}\left\{ \frac{\Theta_{l,m}^{2}(\theta)}{\Theta_{l_{1},m_{1}}^{2}(\theta)}\right\} \right|^{2} 
\sin\theta~\mathrm{d}\theta      
\end{aligned}
\end{equation}

\begin{figure}             
\centering
\begin{minipage}[c]{1.0\textwidth}\centering
\includegraphics[scale=0.90]{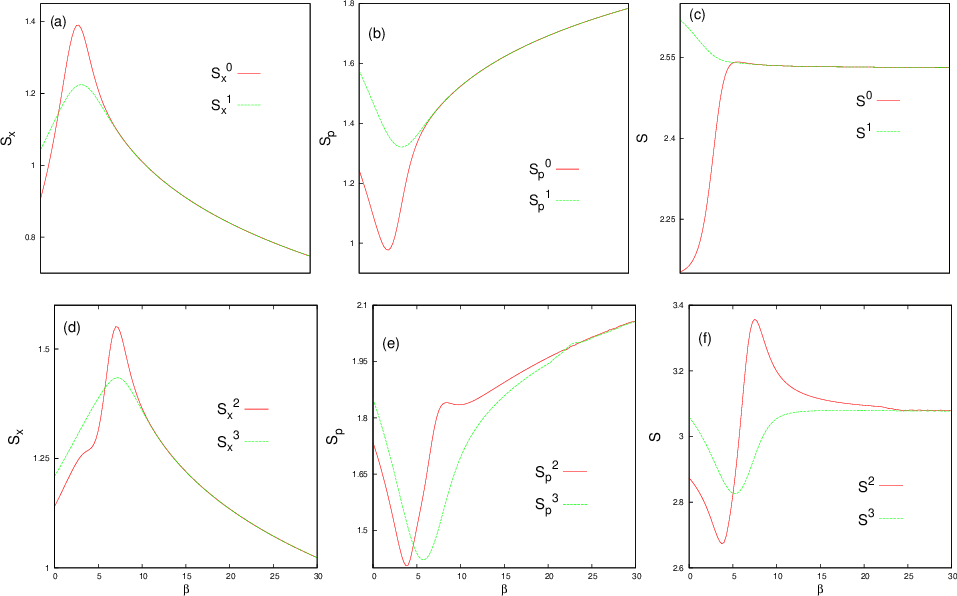}
\end{minipage}%
\caption[optional]{Plots of $S_{x}, S_{p}, S$  vs. $\beta$ \cite{mukherjee15}, at $\alpha=1$, of SDW potential, as in Fig.~\ref{sdw_fish}. }
\label{sdw_shan}
\end{figure}

\section{Result and Discussion}
We will discuss the results in the order they were presented in Sec.~3. Thus first we analyze the information measures in DW potentials, 
followed by CHA and at last the atom-in-a-cage. In all cases, special attention has been paid to the excited states. 

\subsection{Double well potential}
The particular form of DW potential, that has been chosen for the current exposition, has the form: 
$v(x)=\alpha x^{m}-\beta x^{n}+\gamma x^{l}$ ($\alpha, \beta, \gamma$ are positive real numbers, $m,n,l$ are integers and $m>n>l$). Depending
on the values of $m,n, \gamma$, it can fall into one of the two different classes (i) SDW, where $m,n$ are \emph{even} and $\gamma=0$ (ii) 
ADW, where $m,n$ \emph{even}, $\gamma \neq 0$ and $l$ \emph{odd}. Thus it is apparent that, SDW may be regarded as a special case of ADW.

\subsubsection{Symmetric Double well potential}
The main concern is to understand the effect of $\beta$ variation on the behavior of a particle in a SDW potential. An increment in $\beta$ 
enhances both delocalization by increasing spacing between classical turning points, and also assists confinement through increase of barrier 
height. Therefore, an interplay between these two simultaneous and opposing effects should be felt in the pattern of information entropy as 
well--hence one or 
more extremum is to be expected. A detailed study of ground and first excited states reveals that, conventional uncertainty product 
$\Delta x \Delta p$ seems inadequate to explain such features. For both states $\Delta x$ and $\Delta x \Delta p$ advance monotonically 
with rise in $\beta$. On the contrary, $\Delta p$ first decreases, attains a minimum and then progresses with increment of $\beta$. But these 
trends cannot be used to interpret either trapping of the particle or tunneling \cite{mukherjee15}.

\begin{figure}             
\centering
\begin{minipage}[c]{1.0\textwidth}\centering
\includegraphics[scale=0.90]{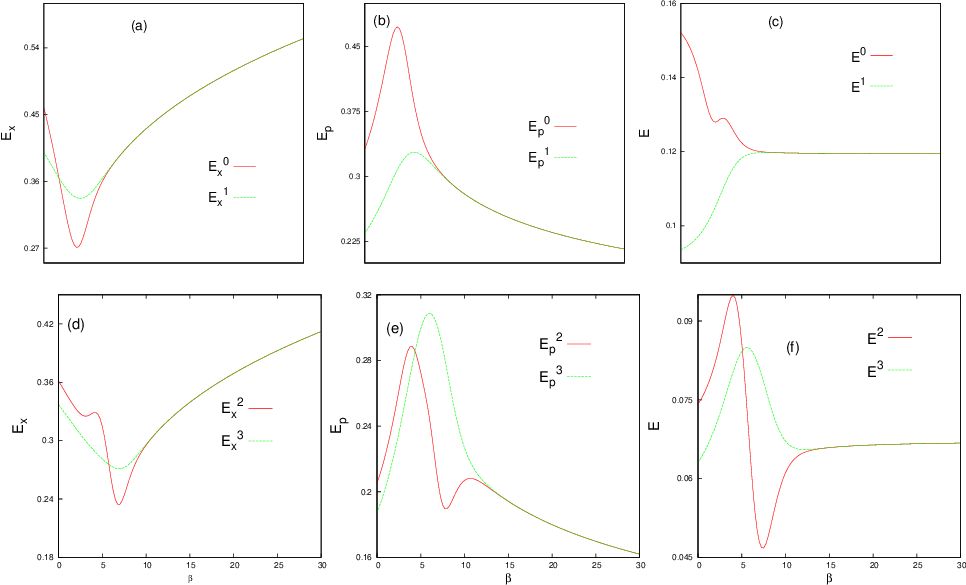}
\end{minipage}%
\caption[optional]{Plots of $E_{x}, E_{p}, E$  vs. $\beta$ \cite{mukherjee15}, at $\alpha=1$, of SDW potential, as in Fig.~\ref{sdw_fish}.}
\label{sdw_onir}
\end{figure}

Against this background, we begin our discussion by first following $I_{x}$, $I_{p}$, $I$. $I_{x}$ for ground and first three excited states 
are qualitatively analogous; 
in all four cases it decreases with increase of $\beta$, attains a minimum and then progressively increases. For a particular state, these minima 
shift to right and gets more flattened with increase of $\alpha$. Appearance of a minimum may be imposed to balance two competing effects. Two 
left panels (a), (d) in Fig.~\ref{sdw_fish} gather plots for $I_{x}^{0}$, $I_{x}^{1}$ and $I_{x}^{2}$, $I_{x}^{3}$ pairs with $\beta$, for a 
definite case of $\alpha=1$. This clearly establishes the merging of two states after a particular $\beta$, which is undoubtedly due to a 
quasi-degeneracy in our SDW potential. This convergence point of $I_x$ shifts to higher values of $\beta$ with rise in $\alpha$.

After observing the extremal nature in $I_x$, it is natural to explore the behavior of $I_p$ or $I$. The qualitative nature of $I_{p}$ and $I$ 
appear to be quite similar. $I_{p}$ progresses with advancement of $\beta$, but the rate of increase gets considerably slower for higher 
$\alpha$. It holds true for all states considered. Also, one finds that, for all these states, at smaller $\beta$, progress in $I_p$ seems 
rather nominal; it remains almost unaltered until a certain $\beta$ is reached, after which $I_p$ keeps on progressing drastically. The 
characteristic $\beta$ at which this transition happens normally increases with state index. Moreover, for a certain state, this $\beta$ is 
shifted to higher values as $\alpha$ advances. Importantly, however, unlike the case of $I_x$, no extremal nature is noticed in this context; 
instead one finds slight flatness for smaller $\beta$. This leads to the inference that, like traditional uncertainty measures, $I_{p}$ also is 
incapable to sense the competitive effects in a SDW. However, as in $I_x$ plots, here also, in two middle panels (b), (e) of Fig.~\ref{sdw_fish}, 
$(I_{p}^{0}$, $I_{p}^{1})$ and $(I_{p}^{2}$, $I_{p}^{3})$ pairs convene at a particular $\beta$ depending on a particular $\alpha$. This also 
could be a probable signature of quasi-degeneracy in this potential. 

Next we present net fisher information $I$. In general, it accelerates with increase in $\beta$ for all four states. Total information initially 
increases very slowly until a certain $\beta$ is reached and after that it imprints drastic continuous growth. For a particular state, the 
$\beta$ at which this transition occurs, is shifted right as $\alpha$ is increased. Progress of $\alpha$ consistently reduces growth rate of $I$ 
in all four occasions. Like the $I_p$ counterpart, $I$ also fails in bearing any characteristic signature suggestive of the competing effects in 
these four states. However, as shown in two rightmost panels (c), (f) of Fig.~\ref{sdw_fish}, pairs like $(I^{0}$, $I^{1})$ 
and $(I^{2}$, $I^{3})$ readily coalesce at a certain $\beta$. On the basis of above exploration, $I$ is not decisive enough to explain 
the dual effects (delocalization and confinement), as well as tunneling.

\begin{figure}             
\centering
\begin{minipage}[c]{1.0\textwidth}\centering
\includegraphics[scale=0.90]{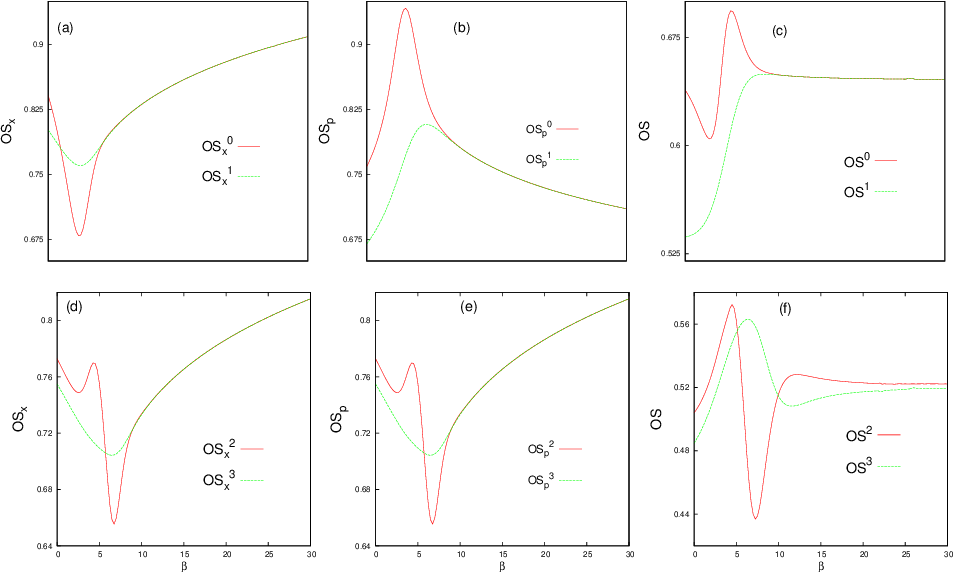}
\end{minipage}%
\caption[optional]{Plots of $OS_x, OS_p, OS$ vs. $\beta$ \cite{mukherjee15}, at $\alpha=1$, of SDW potential, as in Fig.~\ref{sdw_fish}.}
\label{sdw_os}
\end{figure}

Next we focus on $S$ for lowest four states. Apparently, $S_{x}$ at first increases steadily, then attains a maximum and 
finally falls off gradually with increasing $\beta$. Also, for a particular state, with increase of $\alpha$ the positions of these maxima shift 
to higher $\beta$ and respective peak heights decrease. It infers that, at small $\beta$, delocalization effect prevails. However, at large 
$\beta$, the prevalent effect seems to be its confinement. The point of maximum signifies the characteristic $\beta$, at which maximum 
information can be extracted about that state of SDW. It is necessary to mention here that, change in $\alpha$ \emph{does not} cause any 
significant alteration in the qualitative pattern of $S_x$--the maximum shifts to right, and $S_{x}$ values get suppressed at maximum with 
serial increase in $\alpha$. There appears a shoulder in $S_{x}^{2}$; its position also roughly coincides with onset of tunneling. Next, from 
the panel (a) of Fig.~\ref{sdw_shan}, it is noticed that $S_{x}$ for ground and first excited state seem to coalesce at a value of $\beta$ 
approximately close to 5; while same for second and third excited state in panel (d), occurs at closely $\beta=9$. Both these are achieved at 
$\alpha=1$. This merging of $S_{x}$ is a signal of appearance of quasi-degeneracy in both pair of states, and also trapping of particle in 
either of the wells. Study of $S_{p}$ and $S$ will further consolidate this finding.

The general trend of $S_{p}$ for ground and first three excited states seems to be quite similar, \emph{viz.,} $S_{p}$ initially decreases 
sharply with rise in $\beta$, then attains a minimum and continuously increases thereafter. In this case also, variation in $\alpha$ shows no 
qualitative change. Positions of these minima switch to higher $\beta$ with increasing $\alpha$; however this time, values of $S_{p}$ increase 
overall. A shoulder again appears for $S_{p}^2$. However, since effects of tunneling in momentum space is not clearly understood, a definitive 
explanation of this phenomenon appears difficult. In top middle segment (b) of Fig.~\ref{sdw_shan}, $S_p^0$, $S_p^1$ corresponding to first two 
states of SDW potential are plotted against $\beta$, for a fixed $\alpha=1$; analogous plots are drawn for $S_p^2$, $S_p^3$ in bottom middle 
panel (e). Note that, $S_{p}$'s in case of a SDW show a style similar to that found in QHO. While these plots pertain to $\alpha=1$, similar 
trend is recorded for other $\alpha$ as well, with a corresponding shift in location of convergence.

To study the cumulative effect of barrier on a given state, one needs to investigate net $S$. For ground state this quantity 
increases quite sharply, passes through an almost inconspicuous maximum and remains virtually constant thereafter. For first excited state, 
on the other hand, the same decreases monotonically, until individual $\alpha$-plots merge together. Behavior of net $S$ seems to be 
unaffected, at least qualitatively, with respect to variations in $\alpha$--except for a shift in right. The maximum in it (second excited 
state) is much more distinctly defined than its counterpart in ground state. Apart from that, this maximum is now preceded by the appearance 
of a minimum. Now, for third excited state it initially follows same pathway as its counterpart in first excited state, falling steeply and 
monotonically. However, similarities end there; it then attains a minimum, rises rapidly and eventually decays into constancy.

\begingroup
\begin{table}          
\caption{Distribution of lowest six states within two wells, as well as number of effective nodes, for ADW potential, in Eq.~(\ref{pot}). Four 
ranges of $k$ are considered \cite{mukherjee16}.}
\centering
\begin{tabular}{>{\small}c|>{\small}c>{\small}c|>{\small}c>{\small}c|>{\small}c>{\small}c|>{\small}c>{\small}c}
\hline
\hspace{0.1in} $n$ \hspace{0.1in} &    \multicolumn{2}{c}{\hspace{0.2in}$0 \! < \! k \! < \! 1$ } \hspace{0.2in}  
    &    \multicolumn{2}{c}{\hspace{0.2in}$1 \! < \! k \! < \! 2$} \hspace{0.2in}   
    &    \multicolumn{2}{c}{\hspace{0.2in}$2 \! < \! k \! < \! 3$} \hspace{0.2in} 
    &    \multicolumn{2}{c}{\hspace{0.2in}$3 \! < \! k \! < \! 4$} \hspace{0.2in} \\
\hline
    &    well   &   node   &   well    &   node    &  well    &   node    &   well   &  node           \\
\hline
0   &  I    &  0    &  I    &   0   &    I   &   0    &      I    &     0    \\
1   &  II   &  0    &  I    &   1   &    I   &   1    &      I    &     1    \\
2   &  I    &  1    &  II   &   0   &    I   &   2    &      I    &     2    \\
3   &  II   &  1    &  I    &   2   &    II  &   0    &      I    &     3    \\
4   &  I    &  2    &  II   &   1   &    I   &   3    &      II   &     0    \\
5   &  II   &  2    &  I    &   3   &    II  &   1    &      I    &     4    \\
\hline
\end{tabular}
\label{tadw}
\end{table}
\endgroup

Now, two rightmost panels (c), (f) of Fig.~\ref{sdw_shan} illustrate the convergence of net $S$ for ground $(S^0)$, first $(S^1)$ and second 
$(S^2)$, third $(S^3)$ excited states successively, keeping $\alpha$ fixed at 1 in both instances. It becomes stationary at a value of 2.53 on 
and after $\beta=5$ for first pair, while same for latter pair is attained at $\beta=20$ with $S=3.08$. However, these net 
values remain unaltered by variations in $\alpha$. In all cases it obeys the bound given in Eq.~(\ref{eq:20}). 

Leaving aside the second excited state, $E_{x}$ for other three states maintain qualitative agreement amongst themselves, i.e., $E_{x}$ 
initially decreases, attains a minimum, then increases gradually. For a given state, positions of these minima are shifted to higher $\beta$ 
with continuous increase of $\alpha$. Also, the lowest value of $E_{x}$ increases and these minima get flattened with progress in $\alpha$. 
$E_{x}^{2}$ is again different from remaining three states, where one observes a shoulder before reaching minimum and then increases 
monotonically. The position of this shoulder and minima get shifted to right with increase in $\alpha$. It clearly demonstrates the marked 
contrast between second excited state from remaining three states; former shows a sequence of extrema while latter three are characterized by 
a minimum. This uniqueness in $n=2$ could be imposed to the simultaneous effects of $\beta$ on particle--at small values, it indicates the 
increase of delocalization of particle; however at large values, the deciding effect seems to be its confinement. Such kink has been observed 
earlier in case of $S_x^2$. Note that, position of shoulder also roughly coincides with onset of tunneling. Further, from two left segments (a), 
(d) of Fig.~\ref{sdw_onir}, it is transparent that, $E_x$ for lowest two states merge at a value of $\beta \approx 5$ for $\alpha=1$, whereas 
for second, third excited states, convergence occurs at $\beta \approx 9.5$ for same $\alpha$. This joining of $E_{x}$'s is a clear suggestion 
of occurrence of quasi-degeneracy in both pairs of states, and also confinement of particle in either of the wells.

Akin to $E_x$, in $p$ space, the general trend for ground, first and third excited state maintains a qualitative similarity amongst 
themselves; i.e., $E_{p}$ initially advances sharply with increase in $\beta$, reaches a maximum and gradually decreases afterwords. Once again, 
variation in $\alpha$ shows no qualitative change here. Positions of maxima shift to higher $\beta$ and get flatter with rise in $\alpha$. 
variation of $E_{p}^2$ with $\beta$ is, however, substantially different and more interesting; one observes two maxima sandwiched by a minimum 
in $E_{p}^2$. At first, $E_{p}^2$ increases attaining a maximum, then drops down to a minimum, again increases to a maximum and finally 
decreases. Clearly, for progress in $\alpha$, they get right-shifted and individual plots flatten. Again, this effect may be due to 
quasi-degeneracy in this potential. Next, two middle panels (b), (e) of Fig.~\ref{sdw_onir} depict $E_p$s for two lowest pair of 
states, against $\beta$. They meet at nearly $\beta=7.5$ and 11.75. Note that while these plots pertain to $\alpha=1$, 
a similar motif is noticed for other $\alpha$ as well, with a resembling shift in location of convergence point.

\begin{figure}[h]          
\centering
\begin{minipage}[c]{1.0\textwidth}\centering
\includegraphics[scale=0.90]{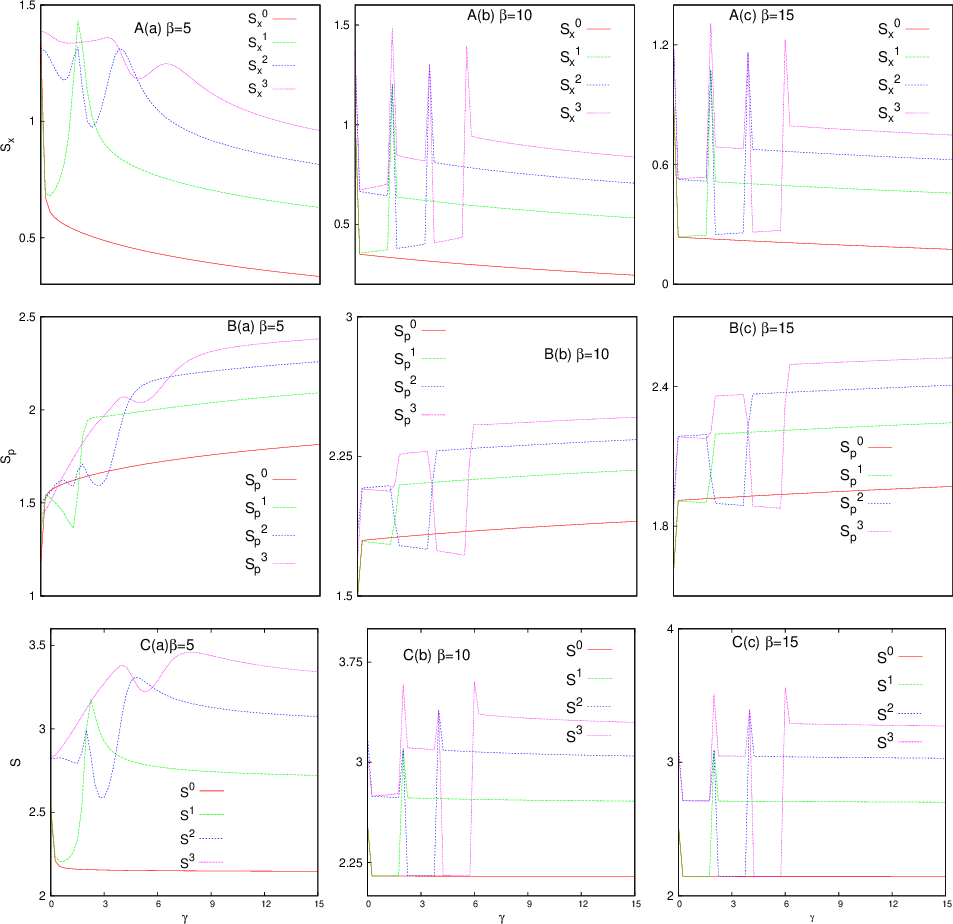}
\end{minipage}%
\caption[optional]{$S_x$ (A), $S_p$ (B), $S_x S_p$ (C) for first four states, in top, middle, bottom rows, plotted against $\gamma$, for 
ADW potential, in Eq.~(\ref{pot}) keeping $\alpha$ fixed at 1. Three panels (a)-(c), in each row refer to three $\beta$, namely, 5,10,15 
respectively \cite{mukherjee16}. For more details, see text.}
\label{adw_shan}
\end{figure}

In case of ground state net $E$ quite sharply decreases initially to attain a minimum, then almost immediately hits a maximum, again reduces 
to reach a constant (0.1195), for all $\alpha$. Locations of both minima and maxima shift towards right for higher $\alpha$. Moreover, 
distance between these two extrema enhances with increase of $\alpha$. In contrast, for first excited state it increases monotonically and 
converge together at 0.1195. This situation modifies for second, third excited states; in former case, it initially rises to reach a maximum, 
then follows through a minimum and again increases until attains a constant (0.0667). The extrema, in this case, are much more distinct and well 
separated than its counterpart in ground state. Values of net $E$ at extrema are quite comparable for all $\alpha$ studied. Increase of $\alpha$ 
transfers the extrema towards right but all of them eventually converge to constant value of 0.0667. For third excited state, it increases to 
reach a maximum, then gradually decreases to attain a constant (0.0667). Now, top right panel (c) of Fig.~\ref{sdw_onir} visualizes convergence of 
ground and first excited state; bottom right panel (f) does same for 2nd and 3rd excited state, keeping $\alpha$ constant at 1 in both occasions. 
Net $E$ becomes stationary at a value of $0.1195$ on and after $\beta=5$ for first pair, while same for latter pair happens at $\beta=20$ with 
net $E=0.0667$. 

Next, we study the Onicescu-Shannon complexity measure (symbolized as OS) for first four states of SDW potential. Except for 
second excited state, $OS_{x}$ decreases with $\beta$, proceeds through a minimum and sharply rises thereafter. Positions of these minima for all 
these three states shift towards right as $\alpha$ is increased. This is again possibly due to competing effects in a SDW. $OS_{x}^{2}$ behaves 
quite differently from remaining three states. Thus, initially there appears a shallow minimum followed by a small maximum and finally a global 
minimum. With $\alpha$, however, positions of these extrema switch towards right as in other three states. Further, segments (a), (d) of 
Fig.~\ref{sdw_os} establish that the pairs $OS_{x}^{0}$, $OS_{x}^{1}$ and $OS_{x}^{2}$, $OS_{x}^{3}$ convene at a certain $\beta$, indicating 
appearance of quasi-degeneracy. In this instance also, qualitative behavior of $OS_{p}$ for ground, first and third excited state are analogous, 
while second excited state stands out. $OS_{p}$ for former three states are characterized by a prominent maximum followed by a gradual decrease, 
while the lone second state exhibits a maximum and minimum in succession. Positions of the maxima switches to higher $\beta$ as $\alpha$ 
increases. As usual, the pairs $OS_{p}^{0}$, $OS_{p}^{1}$ and $OS_{p}^{2}$, $OS_{p}^{3}$ smoothly converge at a particular $\beta$ in middle 
column in (b), (e) of Fig.~\ref{sdw_os}. 
 
\begingroup            
\begin{table}
\small
\caption{ $I_{\rvec}, I_{\pvec}$  for first five $s$ orbitals of CHA, at seven different $r_c$ \cite{mukherjee18d}. See text for details.}
\centering
\begin{tabular}{>{\footnotesize}l|>{\footnotesize}l>{\footnotesize}l>{\footnotesize}l>{\footnotesize}l>{\footnotesize}l>{\footnotesize}l>{\footnotesize}l}
\hline
 $n$ &  $r_c=0.1$ & $r_c=0.5$  & $r_c=1$   &   $r_c=5$ &   $r_c=10$       &    $r_c=20$   &   $r_c=100$     \\
\hline
\multicolumn{8}{c}{$I_{\rvec}^{\dag}$}  \\
\hline
 1   & 3948.73709  & 158.896112  & 40.5850917   & 4.1962752     & 4.00009944  &  4.00000000  & 4.00000000 \\
 2   & 15791.8212  & 632.093249  & 158.322897   & 6.14412880    & 1.48824976  &  1.00148803  & 1.00000000 \\
 3   & 35530.8589  & 1421.49254  & 355.555298   & 14.2086744    & 3.31621534  &  0.70203512  & 0.44444444 \\
 4   & 63165.6652  & 2526.80603  & 631.828614   & 25.2980567    & 6.18599632  &  1.37620355  & 0.25000000 \\
 5   & 98696.1908  & 3947.98150  & 987.090739   & 39.5169227    & 9.79535706  &  2.31004148  & 0.1600061 \\
\hline
\multicolumn{8}{c}{$I_{\pvec}^{\ddag}$}  \\
\hline
 1   & 0.01119745   & 0.26851341     & 1.01251135   &  10.7398867   & 11.9978379    & 11.9999999     &  12.0000000  \\
 2   & 0.01284003   & 0.32261837     & 1.30128804   &  35.6220106   & 114.097280    & 167.397282     &  168.000000 \\
 3   & 0.01312184   & 0.32950852     & 1.32621209   &  35.4490641   & 152.667868    & 568.122689     &  827.999999 \\
 4   & 0.01321715   & 0.33151636     & 1.33186010   &  34.7398719   & 146.840310    & 640.181092     &  2591.99999  \\
 5   & 0.01326029   & 0.33232931     & 1.33360252   &  34.3394100   & 142.869003    & 610.819403     &  6299.58443 \\
\hline
\end{tabular}       
\begin{tabbing}
$^\dag$$I_\rvec$ of FHA for $n=1-5$ are: 4,~1,~$\frac{4}{9}$,~0.25,~0.16. \\
$^\ddag$$I_\pvec$ of FHA for $n=1-5$ are: 12,~168,~828,~2592,~6300. \\
\end{tabbing}
\label{fish_t}
\end{table}
\endgroup
$OS^{0}$  at first, falls down to a minimum slowly, then grows to a maximum very sharply, eventually becoming constant. These extrema shift 
towards higher $\beta$ with increasing $\alpha$; however, they finally converge to same constant (0.6465). $OS^{1}$ initially increases 
with $\beta$, then converges to a stationary value of 0.6465. Rate of increase of $OS^{1}$ with $\beta$ declines with increase of $\alpha$. 
$OS^{2}$ initially rises to a maximum, then follows through a minimum and again convenes to a constant (0.525). All these extrema move toward 
right with increase of $\alpha$. $OS^{3}$ rises to a maximum and converges to same value (0.525). In this case also, a progress in $\alpha$ 
leads to same result of shifting maxima towards right. Additionally, in two rightmost panels (c), (f) of Fig.~\ref{sdw_os}, we provide
convergence of $OS$ for first and second pairs respectively. More elaborate works of information measure of SDW is available in 
\cite{mukherjee15}.

\subsubsection{Asymmetric Double well potential}
An ADW potential is invoked by adding a linear term in SDW potential. This asymmetric term ($\gamma x$) immediately lifts the quasi-degeneracy 
in energy. However, a new kind of degeneracy emerges at certain characteristic $\gamma$. Extensive study of energy exposes that, at each 
$\alpha$ there appears a range of $\gamma$ after which energy increases steadily. It has been found that, for a fixed $\alpha$, there exists a 
\emph{positive} real number $k=\frac{\gamma}{\Delta \gamma}$, depending upon which, following rules for quasi-degeneracy in energy has been 
proposed \cite{mukherjee16}.   

\begin{enumerate}[(i)]      
\item 
When $(n \! \geq \! k)$: three possible outcomes can be envisaged in this scenario.

\begin{enumerate}[(a)]
\item
\emph{$k$ is odd positive integer}: an \emph{odd}-$n$ state will be quasi-degenerate with its immediate higher state.
\item
\emph{$k$ is even positive integer}: an \emph{even}-$n$ state will be quasi-degenerate with its adjacent higher state.
\item
\emph{$k$ is a fraction}: no quasi-degeneracy is possible.                   
\end{enumerate}
\item
 When $k \! > \! n$: $n$-th state \emph{can not} be quasi-degenerate. However, other higher ($\! > \! n$) states may 
show degeneracy depending on $k$, as delineated above in (i.a), (i.b). 
\end{enumerate}

These quasi-degeneracy rules were demonstrated in \cite{mukherjee16}. A in-depth analysis also tells that, at a definite $\alpha$, 
by controlling $\beta, \gamma$ desired quasi-degenerate pair can be generated. Examination of wave functions and densities for various 
states \cite{mukherjee16} uncovers that, a particle in ground state always 
resides inside deeper well. On the other hand, particle in an arbitrary excited state oscillates between deeper and shallower wells, and 
eventually gets trapped inside deeper well. This observation has been quantified in terms of some simple rules to forecast 
localization/delocalization of a particle in a certain well, in terms of the parameter $k$$(=\frac{\gamma}{\Delta \gamma})$ and state index 
$n$. For $\mathcal{E}_{n} < V_{0}$, following situations could be visualized.

\begin{enumerate}[(i)]
\item
$n \! \geq \! k$: two possibilities arise:
\begin{enumerate}[(a)]
\item
$k$ is integer: particle in $n$th state is distributed in both well~I and well~II.
\item
$k$ is fraction: four possibilities need to be considered: 

\begin{enumerate}[(1)]
\item
both $n$ and integer part of $k$ $even$: particle stays in well~I.
\item
both $n$ and integer part of $k$ $odd$: particle stays in well~I.
\item
$n$ $even$ and integer part of $k$ $odd$: particle stays in well~II.
\item
$n$ $odd$ and integer part of $k$ $even$: particle stays in well~II.
\end{enumerate}
\end{enumerate}
\item
$n \! < \! k$: particle always resides in larger (I) well.
\end{enumerate}

\begin{figure}                         
\begin{minipage}[c]{1.0\textwidth}\centering
\includegraphics[scale=0.90]{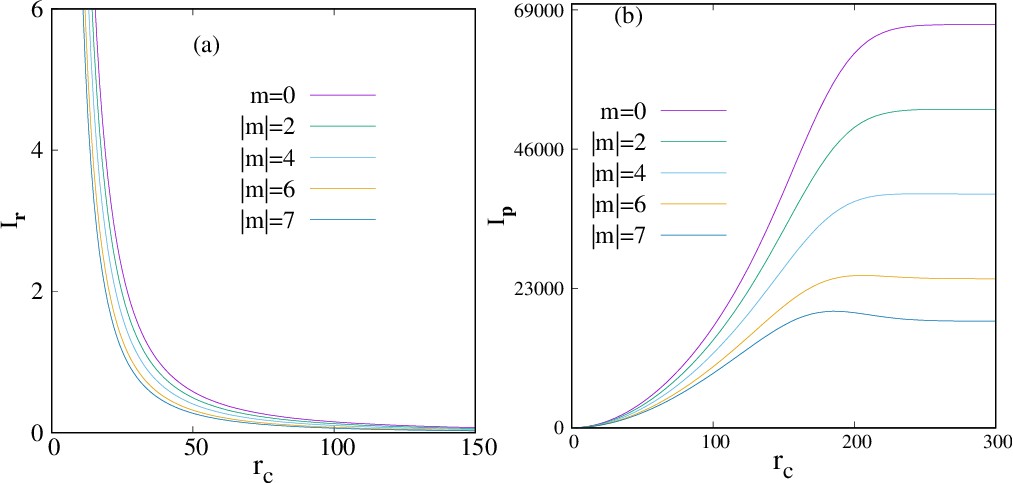}
\end{minipage}%
\caption{Variation of $I_{\rvec}, I_{\pvec}$ in CHA, with $r_c$, at five selected $|m|$ ($0, 2, 4, 6, 7$) of 
$10k$ orbital, in panels (a), (b) respectively \cite{mukherjee18d}. See text for details.}
\label{fisr_h_1}
\end{figure}

It is worthwhile making some remarks about nodes in wave functions. Table~\ref{tadw} provides number of effective nodes present in a particular 
state, at same selected range of $k$. In fact, this increases with state index ($n$th state has $n$ nodes lying within classical turning 
points). However, because of occurrence of two separate wells as discussed above, certain nodes become insignificant. In range 
$0 \! < k \! < 1$, for example, $n \! = \! 0$,1 are practically node-less as they behave as ground state of wells~I, II respectively. 
Analogously, $n \! = \! 2$,3 possess single node, for they represent first excited state of wells~I, II; whereas $n \! = \! 4$,5 contain two 
nodes corresponding to second excited states of wells~I, II respectively. These numbers are provided in third column of this table. Next, fifth 
column concludes that, within $1 \! < \! k \! < 2$, $n \! = \! 0$,2 have zero node; thus they represent the lowest two states of wells~I, II. 
Also, $n \! = \! 1$,4 have single node, as they appear as first excited states of wells~I, II. And $n \! = \! 3$,5 possess 2,3 nodes 
respectively, for they relate to second and third excited states of well~I. In a similar fashion, in range $2 \! < k \! < \! 3$, 
$n \! = \! 0$,1,2,4 behave as lowest four states of well~I, whereas $n \! = \! 3$,5 act as lowest two states of well~II. These are clear from 
column seven. Finally for range $ 3 \! < \! k \! < \! 4$, last column ensures that $n \! = \! 0$,1,2,3,5 serve as first five states of well~I, 
giving rise to 0,1,2,3,4 nodes respectively, whereas $n \! = \! 4$ remains effectively node-less, as it becomes the ground state of well~II. 
Thus, one concludes that at fractional $k$, wells~I and II virtually function as two different potentials (after a certain threshold $\beta$).

We now move on to information measures for first four energy states. Figure~\ref{adw_shan} imprints $S_{x}$ as a function of $\gamma$ at 
three $\beta$ ($5,10,15$) in top three rows, A(a)-A(c). Increase in $\beta$ only makes the particle more trapped within well~I or II. Generally, 
at certain $\gamma$, plot for $n$th state quickly jumps to a crest and then gradually decreases by traveling through $n+1$ peaks. After that, 
particle ultimately localizes in larger (I) well. These jumps at even $\gamma$ (integer $k$) convey transition points, which occurs only at 
fixed interval of $\Delta \gamma$, and observed up to a certain $\gamma$ (characteristics for a given state). At fractional $k$, development in 
$S_{x}$ is rather small because of confinement. Once, it finally rests in larger well ($k \! > \! n$), $S_{x}$ decreases with rise in $\gamma$ 
explaining gradual build-up of confinement. Next B(a)-B(c), delineate that slope of $S_{p}$ against $\gamma$ curves varies noticeably at 
intervals of $\Delta \gamma$. This discontinuity indicates the transition of particle from one well to another. Again, an $n$th state leads 
to ($n$+1) jumps in $S_{p}$ before localization occurs in well~I. Progress of net $S$ as a function of $\gamma$ are shown in 
bottom row. Like $S_{x}$, net $S$ also portrays pronounced peaks at intervals of $\Delta \gamma$ indicating transition points at integer $k$. 
Apart from that, like $S_x$, an $n$th state gives rise to ($n$+1) peaks in $S_p$ before settling in well~I. When particle resides in a specific 
well, extent of confinement is not always same. Visibly, when $0 \! < \! k \! < \! 1$ ($ 0 \! < \! \gamma \! < \! 2 $), net $S$ for 
$n \! = \! 0,1$ remain very intimate. This is narrated from the fact that, at this interval these two states act effectively as lowest state 
of wells~I, II. Likewise, for $1 \! < \! k \! < \! 2$ ($ 2 \! < \! \gamma \! < \! 4$), net $S$ of $n \! = \! 0,2$ are closer to each other, as 
they behave as lowest states of wells~I, II. As anticipated, in $2 \! < \! k \! < \! 3$, net $S$ of $n \! = \! 0,3$ approach each other quite 
closely. This observations could be explained from a consideration of number of effective nodes in Table~\ref{tadw} in each of these states in 
respective interval of $\gamma$. Thus, at $n$th interval of $\Delta \gamma$, $n$th state behaves similar to ground state of a definite well. 
Note that there is a jump in $S_{x}^{0},S_{p}^{0}, S^{0}$ at $\gamma \! = \! 0$, informing a critical transition point, after which a particle 
remains in well~I. This also mirrors the reality that, SDW is a special case of ADW potential.

\begin{figure}                         
\begin{minipage}[c]{1.0\textwidth}\centering
\includegraphics[scale=0.90]{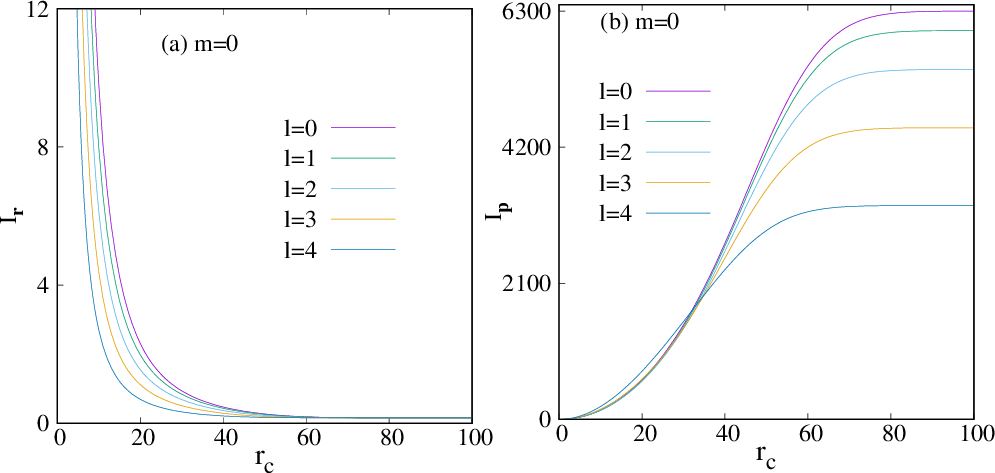}
\end{minipage}%
\caption{Variation of $I_{\rvec},I_{\pvec}$ in CHA, with $r_c$, for $5s$-$5g$ states keeping $|m|=0$, in panels 
(a), (b) respectively \cite{mukherjee18d}. See text for details.}
\label{fishr_h_2}
\end{figure}

We next examine progress of $S$ with $\beta$ through four lowest states taking $\gamma$ as parameter. 
In all cases $S_{x}$ initially rises with $\beta$ before attaining a maximum and then declines asymptotically. This arises 
due to the dual effect of $\beta$. Interestingly, positions of these maxima move to left with increase in $\gamma$, for any given state; 
moreover they tend to vanish with $\gamma$, with higher state requiring larger $\gamma$. This is illustrated by considering the predominance 
of asymmetry over competing effect due to localization of particle in a definite well. Depending upon odd, even values of $\gamma$, two different 
tendency in $S_{x}$ is seen beyond a certain threshold $\beta$. For $k \! = \! 0$ ($\gamma \! = \! 0$), $S_{x}$ of $n \! = \! 0$,1 are very close; 
similarly at $k \! = \! 1$ ($\gamma \! = \! 2$), same happens for $n \! = \! 1$,2; at $k \! = \! 2$ ($\gamma \! = \! 4$), $n \! = \! 2$,3 are 
close. Continuing in this fashion, one can predict that, $S_{x}$ of $n \! = \! 3$,4 will approach each other at $k \! = \! 3$ or 
$\gamma \! = \! 6$ (not shown) and so on. This closeness of various sets of $S_{x}$ at certain $k$ is in consonance with quasi-degeneracy rule. 
On the other hand, when $\gamma$ is odd (1, 3, 5, $\cdots$), then at $k \! = \! 0.5, 1.5, 2.5$, $S_{x}$ of $n \! = \! 0$ becomes closer to 
$n \! = \! 1$,2,3 states respectively. This, again is due to number of effective node of those states becoming equal at those specific $k$.   

One notices that, generally, $S_{p}$ declines with rise in $\beta$, attains a minimum and then increases gradually. Again, positions of these 
minima move towards left with progress in $\beta$. Beyond a particular $\gamma$, the extrema tend to die out. Also for higher $n$, there exist
some occasional humps in $S_{p}$. Alike to $S_x$, $S_p$ in certain states also converge depending on odd, even $\gamma$. However, the alterations 
of net $S$ with $\beta$ is not straightforward; complicated sequence of maxima, minima is observed for different states for various $\beta$. 

\begingroup           
\begin{table}
\small
\caption{$R_{\rvec}^{\alpha}, R_{\pvec}^{\beta}$ and $R^{(\alpha, \beta)}= (R_{\rvec}^{\alpha}+R_{\pvec}^{\beta})$ for lowest 
two $s$ states in a CHA, at various $r_c$, for $\alpha=\frac{3}{5}, \beta=3$ respectively. See text for more details.}
\centering
\begin{tabular}{>{\footnotesize}l>{\footnotesize}l>{\footnotesize}l>{\footnotesize}l|>{\footnotesize}l>{\footnotesize}l>{\footnotesize}l>{\footnotesize}l}
\hline
\multicolumn{4}{c}{$1s$}   \vline &      \multicolumn{4}{c}{$2s$}    \\
\cline{1-4}  \cline{5-8}
$r_c$  &    $R_{\rvec}^{\alpha}$     & $R_{\pvec}^{\beta}$  &  $R^{(\alpha, \beta)}$  &  
$r_c$  &    $R_{\rvec}^{\alpha}$     & $R_{\pvec}^{\beta}$  &  $R^{(\alpha, \beta)}$  \\
\hline 
0.1   & $-$6.04495302  & 12.254494   &  6.209541  &  0.1   &  $-$6.06527856  &   14.246181   & 8.180902      \\
0.2   & $-$3.97406865  & 10.182673   &  6.208604  &  0.2   &  $-$3.98570108  &   12.160542   & 8.174841      \\
0.5   & $-$1.25276392  &  7.458575   &  6.205811  &  0.5   &  $-$1.23590502  &    9.387514   & 8.151609      \\
1.0   &    0.77359587  &  5.427664   &  6.201260  &  1.0   &     0.84696859  &    7.245474   & 8.092443      \\                                                  
5.0   &    4.66206639  &  1.524658   &  6.186724  &  5.0   &     5.76615415  &    1.215531   & 6.981685     \\
7.5   &    4.93915225  &  1.263505   &  6.202657  &  7.5   &     6.96559616  & $-$0.391676   & 6.573919    \\
10.0  &    4.97268114  &  1.238720   &  6.211402  &  10.0  &     7.70531072  & $-$1.298410   & 6.406900     \\
40.0  &    4.97592206  &  1.237321   &  6.213243  &  40.0  &     8.61299696  & $-$2.335383   & 6.277613    \\
\hline
\end{tabular}
\label{renyi_table_cha}
\end{table}
\endgroup
Similar study was performed for development of $I, E$, as well as Onicescu-Shannon information measures with changes in $\gamma$ and $\beta$. 
A careful analysis unveils that, nature of $I_{x},E_{x}$, $OS_{x}$ with increase of $\gamma$ are qualitatively similar to those of $S_{p}$. 
Likewise, trends of $I_{p},E_{p}$, $OS_{p}$ with rise in $\gamma$ resemble the behavior of $S_{x}$. The conclusions from $\beta$ variation are 
also in harmony to $S$. Overall all these measures lead to comparable conclusions as obtained through conventional uncertainty product, 
and $S$ in position and momentum space.

\subsection{Hydrogen atom}
\subsubsection{Information analysis for CHA}
At the beginning it is appropriate to point out that, in case of central potential \emph{net} information measures in both $r$ and $p$ spaces 
are separated into radial and angular parts. In a given space, the results provided resemble that of net measure including angular part. An FHA 
can be transformed to CHA by shifting the boundary from \emph{infinity} to finite region. This variation in radial environment does not 
influence the angular boundary conditions. Hence, the angular contribution of information measure in FHA and CHA remain unaltered in both $r$ 
and $p$ spaces. For calculation of $S, R, E$ magnetic quantum number $m$ has been kept fixed to $0$. However, $I$ has been investigated for 
non-zero $m$ states. The radial wave functions in $r$ and $p$ spaces depend only on $n, l$ quantum number. Hence, in both spaces radial wave 
function can be achieved by taking $m=0$. Apart from that, a change in $m$ from \emph{zero} to \emph{non-zero} value does not affect the 
expression of radial wave function in $p$ space. 

At first, $I_{\rvec}, I_{\pvec}$ are computed from Eq.~(\ref{fish_cp}). In this equation, angular part is normalized to unity. Hence, estimation 
of these derived quantities by employing only radial part serves the purpose. In order to understand the effect of confinement using $I$, pilot 
calculations are done by the authors in \cite{mukherjee18d}, for $1s,2s,2p,3s,3d,4s,4f$ and all $l$-states corresponding to $n=5$ and 10, varying 
$r_{c}$ from 0.1 to 300 a.u. This chapter discusses some of these results.
    
\begin{figure}                                            
\begin{minipage}[c]{1.0\textwidth}\centering
\includegraphics[scale=0.90]{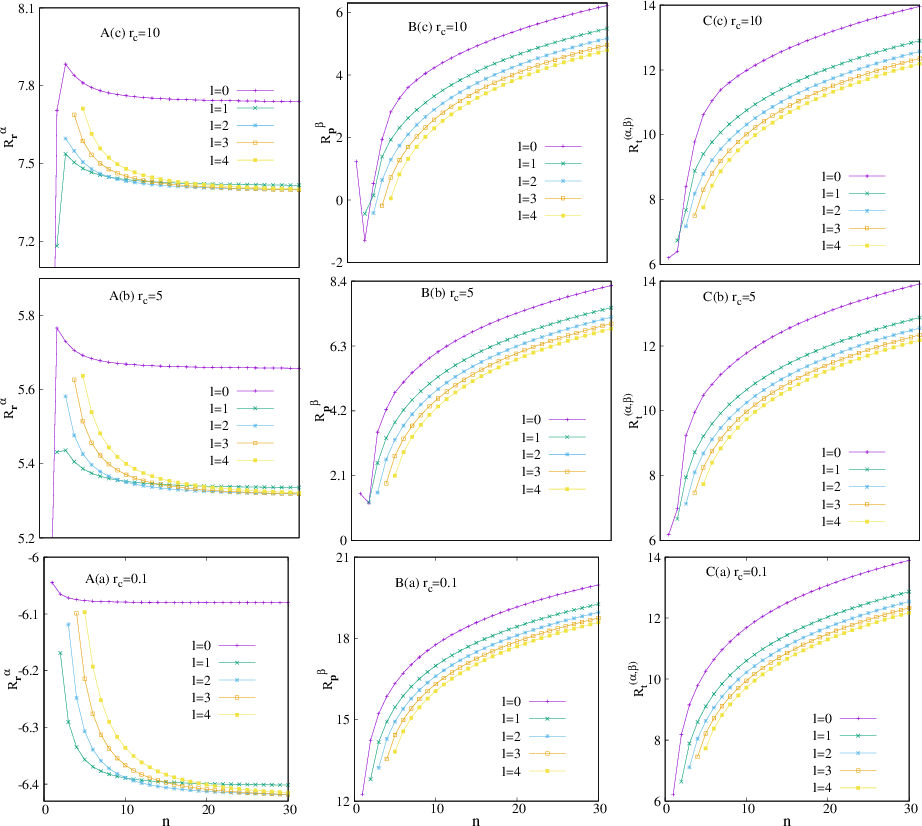}
\end{minipage}%
\caption{Plot of $R^{\alpha}_{\rvec}$ (A), $R^{\beta}_{\pvec}$ (B) and $R_{t}^{(\alpha, \beta)}$ (C) versus $n$ for $s,p,d,f,g$ 
orbitals at three particular $r_{c}$ values of CHA, namely, $0.1,5,10$ in panels (a)-(c). $R_{t}^{(\alpha,\beta)}$'s 
for all these states obey the lower bound given in Eq.~(\ref{eq:21}). For more details, consult text.}
\label{renyi_n_cha}
\end{figure}

As discussed before, $I$ is a measure of fluctuation in a given probability distribution. The analytical forms of $I_{\rvec}$ and $I_{\pvec}$ 
in a FHA were given in \cite{romera05},
\begin{eqnarray} \label{fish_fha}
I_{\rvec}(Z) & = & \frac{4Z^2}{n^2}\left[1-\frac{|m|}{n}\right], \\
I_{\pvec}(Z) & = & \frac{2n^2}{Z^2} \left[ \left(5n^2+1-3l(l+1)\right)-|m|(8n-6l-3) \right]. 
\end{eqnarray}
Equation~(\ref{fish_fha}) endorses that, at a certain $m$, $I_{\rvec}(Z)$ declines with progress in $n$ and for a given $n$, linearly decreases 
with growth in $m$. However, it remains unaffected with change in $l$. At $m=0$, $I_{\rvec}=4\langle p^{2} \rangle =8 \times KE$ ($KE$ is the average 
kinetic energy). On the other hand, at a definite $l,m$, $I_{\pvec}(Z)$ rises with $n$; but decreases with advancement of $l$ for fixed $n,m$ 
values. Similarly, for a certain $n,l$, $I_{\pvec}(Z)$ regresses with enhancement of $|m|$.

A careful analysis suggests that the trends of $I_{\rvec},I_{\pvec}$ in CHA are not always in consonance with FHA. Table~\ref{fish_t} 
imprints the behavior of $I_{\rvec}, I_{\pvec}$ for first five $s$ orbitals of CHA at seven different $r_{c}$ ($0.1, 0.5,1,5,10,20,100$). 
$I_{\rvec}$ for a definite $n,l,m$ state declines with enhancement in $r_{c}$. In FHA, $KE=\frac{Z^{2}}{n^{2}}$ decreases with rise 
in $n$. On the contrary, in case of CHA, at a fixed $r_c$, $KE$ increases with progress in $n$. This trend modifies at higher $r_c$. 
$I_{\pvec}$ at a fixed $n,l,m$ state enhances with growth in $r_c$. However, unlike $I_{\rvec},$ pattern of $I_{\pvec}$ with change of $n$ is 
straightforward. At low $r_c$ region, it decreases with $n$ (at fixed $l,m$). For all these states, $I_{\rvec}, I_{\pvec}$ eventually converge to 
respective FHA limits. 

\begingroup           
\begin{table}
\small
\caption{$S_{\rvec}, S_{\pvec}$,  $S$ for $1s$, $2s$ states in CHA at some chosen $r_c$. $S$'s 
for all these states obey the lower bound given in Eq.~(\ref{eq:20}).  See text for details.}
\centering
\begin{tabular}{>{\footnotesize}l>{\footnotesize}l>{\footnotesize}l>{\footnotesize}l|>{\footnotesize}l>{\footnotesize}l>{\footnotesize}l>{\footnotesize}l}
\hline
\multicolumn{4}{c}{$1s$}   \vline &      \multicolumn{4}{c}{$2s$}    \\
\cline{1-4} \cline{5-8}
$r_c$  &    $S_{\rvec}$    & $S_{\pvec}$  &  $S=S_{\rvec}+S_{\pvec}$          &
$r_c$  &    $S_{\rvec}$    & $S_{\pvec}$  &  $S=S_{\rvec}+S_{\pvec}$          \\
\hline 
0.1  & $-$6.24450338  & 12.8535    &  6.6089    &   0.1   &  $-$6.44745791     &  14.638      &    8.1905    \\
0.2  & $-$4.17785640  & 10.7787    &  6.6008	&   0.2   &  $-$4.36923353     &  12.5593     &    8.1900    \\
0.5  & $-$1.47034068  &  8.0472    &  6.5768    &   0.5   &  $-$1.62307869     &  9.8112      &    8.1881    \\
1.0  &    0.52903030  &  6.0114    &  6.5404    &   1.0   &     0.45546229     &  7.7347      &    8.1901    \\
5.0  &    4.01744418  &  2.524361  &  6.541805  &   5.0   &     5.46416082     &  2.8173      &    8.2814    \\
7.5  &    4.13932453  &  2.425508  &  6.564832  &   7.5   &     6.72302624     &  1.3022      &    8.025    \\
10.0 &    4.14460143  &  2.421936  &  6.566538  &   10.0  &     7.44615626     &  0.2765      &    7.7226    \\
40.0 &    4.14472988  &  2.421862  &  6.566592  &   40.0  &     8.11092936     &  $-$0.75758  &    7.35334    \\
\hline
\end{tabular}
\label{shan_table_cha}
\end{table}
\endgroup

At fixed $n,l,r_{c}$, the effect of $m$ quantum number on $I_{\rvec}, I_{\pvec}$ is candid. Like FHA, both $I_{\rvec}, I_{\pvec}$ in CHA decrease 
with increment in $|m|$. A sample result for $10k$ state is given in panels (a) and (b) of Fig.~\ref{fisr_h_1} for $I_{\rvec},I_{\pvec}$ 
respectively, as a function of $r_{c}$ at five different $|m|$, namely $0,2,4,6,7$. It is interesting to mention that, $I_{\pvec}$ for 
$l \neq 0$ states having $m=0$, increases with progress in $r_{c}$ and finally reaches to their FHA values. However, for \emph{non-zero} $l,m$ 
(at a fixed $n$) states, it passes through a maximum before reaching FHA limit.

It is well-known that, $I_{\rvec}$ in FHA is independent of $l$ quantum number. However, in CHA, at a certain $n,m,r_{c}$ it decreases 
with increment of $l$. One such example is panel (a) of Fig.~\ref{fishr_h_2}, where, $I_{\rvec}$ is plotted against $r_c$ for $5s-5g$. 
This graph complements above conclusion about $I_{\rvec}$; at low $r_c$ region it reduces with $l$ and at $r_{c} \rightarrow \infty$, becomes 
invariant of $l$. At a certain $n,|m|,r_{c}$, $I_{\pvec}$ in CHA advances with rise in $l$. However, in FHA it exerts a complete opposite trend,
reducing with growth in $l$. Hence, there exists a crossover region in $r_{c}$ after which the behavioral pattern of $I_{\pvec}$ alters to match 
the respective FHA. Panel (b) of Fig.~\ref{fishr_h_2} establishes this, where $I_{\pvec}$ for $5s-5g$ are pictorially given as 
functions of $r_c$.             

The authors have investigated $S,R,E$ in both $r,p$ spaces for various states of CHA \cite{mukherjee18c,mukherjee18e}. Here we take this opportunity
to address some of the interesting outcomes of the above exploration. Let us begin with $R$. In case of all $n,l,m$ states 
of CHA, $R_{\rvec}^{\alpha}$ increases with rise in $r_c$. Conversely, $R_{\pvec}^{\beta}$ for all these states declines with growth in 
$r_c$. Table~\ref{renyi_table_cha} imprints the values of $R_{\rvec}^{\alpha}, R_{\pvec}^{\beta}$ and $R^{(\alpha, \beta)}$ for $1s$ and 
$2s$ states at eight selected $r_c$ values. Apart from that, it is important to mention that, at \emph{small} $r_c$, R\'enyi
entropies for $2p,3d,4f,5g$ states in $r, p$ spaces obey same order: $R_{\rvec}^{\alpha}(5g)>R_{\rvec}^{\alpha}(4f)>R_{\rvec}^{\alpha}(3d) 
>R_{\rvec}^{\alpha}(2p)$ and $R_{\pvec}^{\beta}(5g)>R_{\pvec}^{\beta}(4f)>R_{\pvec}^{\beta}(3d)>R_{\pvec}^{\beta}(2p)$. 
But at $r_c \rightarrow \infty$ limit, ordering in $p$ space completely reverses, while the $r$-space ordering is preserved. 

\begin{figure}                                            
\begin{minipage}[c]{1.0\textwidth}\centering
\includegraphics[scale=0.90]{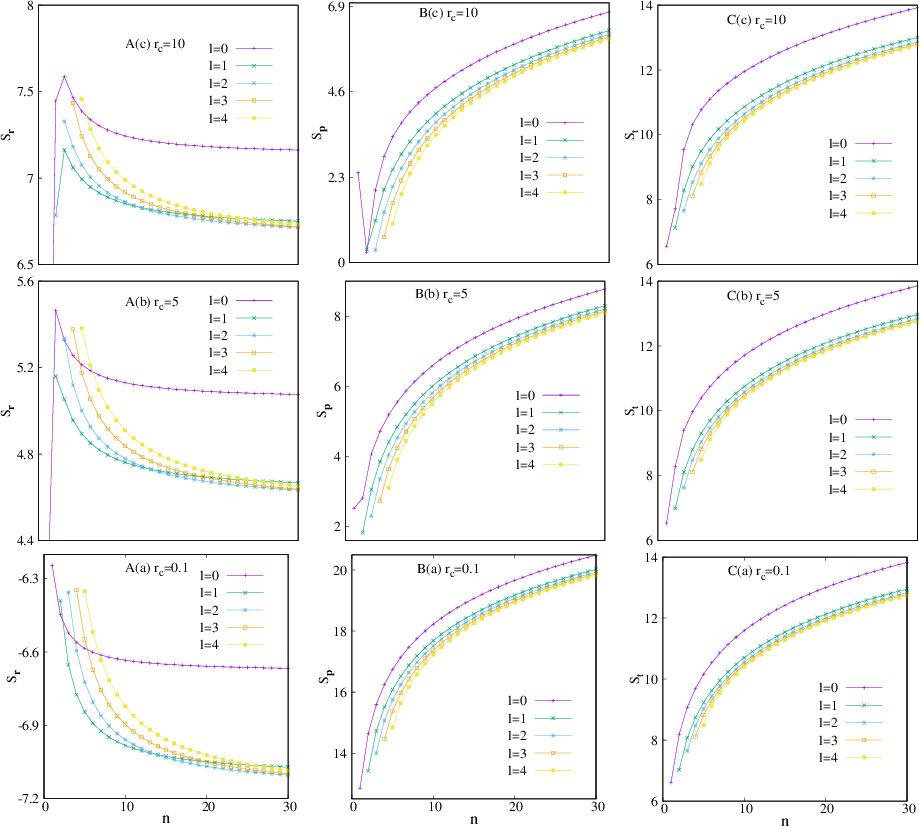}
\end{minipage}%
\caption{Plot of $S_{\rvec}$ (A), $S_{\pvec}$ (B) and $S_{t}$ (C) versus $n$ for $s,p,d,f,g$ 
orbitals at three particular $r_{c}$ values of CHA, namely, $0.1,5,10$ in panels (a)-(c). $S_{t}$'s 
for all these states obey the lower bound given in Eq.~(\ref{eq:20}). For more details, consult text.}
\label{shan_n_cha}
\end{figure}

Dependence of $R$ on $n$ is quite interesting. At a constant $l$ and $r_{c}=0.1$, $R_{\rvec}^{\alpha}$ declines with
rise in $n$, while $R_{\pvec}^{\beta}$ and $R_{t}^{\alpha,\beta}$ show reverse tendency. This has been pictorially represented in
bottom panels A(a), B(a), C(a) of Fig.~\ref{renyi_n_cha} respectively. This clearly explains that, effect of confinement enhances
with progress in $n$. In this regard, it is worthwhile mentioning that, at $r_{c} \rightarrow \infty$, exactly opposite pattern with 
respect to $R_{\rvec}^{\alpha}$ and $R_{\pvec}^{\beta}$ is seen. First two columns of middle and top rows of Fig.~\ref{renyi_n_cha}     
indicate the appearance of maxima and minima in the plots in $r,p$ spaces respectively. This suggests that with increase in $r_c$,  
the system approaches towards FHA. These extrema will shift to right with rise in $r_{c}$. Now, regarding the dependence of $R$ on $l$, at a 
fixed $n$, $R_{\rvec}^{\alpha}$ initially decreases to attain a minimum and then increases with rise in $l$. However, $R_{\pvec}^{\beta}$ under 
same condition (fixed $n$), decreases with progress in $l$. This has been thoroughly discussed in \cite{mukherjee18e} for $n=10$ states. 

Now we move on to $S$. Akin to $R_{\rvec}^{\alpha}$ and $R_{\pvec}^{\beta}$, $S_{\rvec}$ and $S_{\pvec}$ for a given $n,l,m$ state progresses 
and declines with rise in $r_c$. Table~\ref{shan_table_cha} displays $S_{\rvec}, S_{\pvec}, S_{t}$ for $1s$ and $2s$ states at eight selected 
$r_c$. Moreover, $S_{\rvec}$ imprints separate pattern to $R_{\rvec}^{\alpha}$ but $S_{\pvec}$ depicts similar behavior favouring to 
$R_{\pvec}^{\beta}$ in smaller $r_c$. The observed nature for $2p,3d,4f,5g$ in $r, p$ spaces is slightly unexpected: 
$S_{\rvec}(4f)>S_{\rvec}(5g)>S_{\rvec}(3d)>S_{\rvec}(2p)$ and $S_{\pvec}(5g)>S_{\pvec}(4f)>S_{\pvec}(3d)>S_{\pvec}(2p)$ \cite{mukherjee18e}. 
As seen before, at $r_c \rightarrow \infty$ the order in $r, p$ spaces reorganize to $S_{\rvec}(5g)>S_{\rvec}(4f)>S_{\rvec}(3d)>S_{\rvec}(2p)$ and 
$S_{\pvec}(5g)<S_{\pvec}(4f)<S_{\pvec}(3d)<S_{\pvec}(2p)$. 

Our next objective is to understand the dependence of $S$ on $n$. At a constant $l$ and $r_{c}=0.1$, $S_{\rvec}$, $S_{\pvec}$  and $S_{t}$ resemble 
that of $R_{\rvec}^{\alpha}$, $R_{\pvec}^{\beta}$ and $R_{t}^{\alpha,\beta}$ respectively. 
The former decreases and latter two increase with rise in $n$. This has been exhibited in bottom panels A(a), B(a), C(a) of 
Fig.~\ref{shan_n_cha} successively. This again explains the fact that, effect of confinement accelerates
with advancement in $n$. It is needless to mention that, at $r_{c} \rightarrow \infty$, exactly reverse manner for $S_{\rvec}$ and $S_{\pvec}$ 
is observed. Here again, first two columns of middle and top rows of Fig.~\ref{shan_n_cha} suggest the appearance of maxima and minima in 
the plots in $r,p$ spaces respectively. This suggests that, with relaxation in confinement the system approaches towards FHA. These extrema 
will move towards right with rise in $r_{c}$. We can now follow the effect of $l$ on $S$. Similar to $R_{\rvec}^{\alpha}$, here 
also at a constant $n$, $S_{\rvec}$ initially falls down to a minimum and then increases with progress in $l$. Parallel to $R_{\pvec}^{\beta}$, 
$S_{\pvec}$ at a fixed $n$, declines with growth in $l$. This has been discussed in detail \cite{mukherjee18e} using $n=10$ states as reference. 

\begin{figure}[h]                                            
\begin{minipage}[c]{1.0\textwidth}\centering
\includegraphics[scale=0.90]{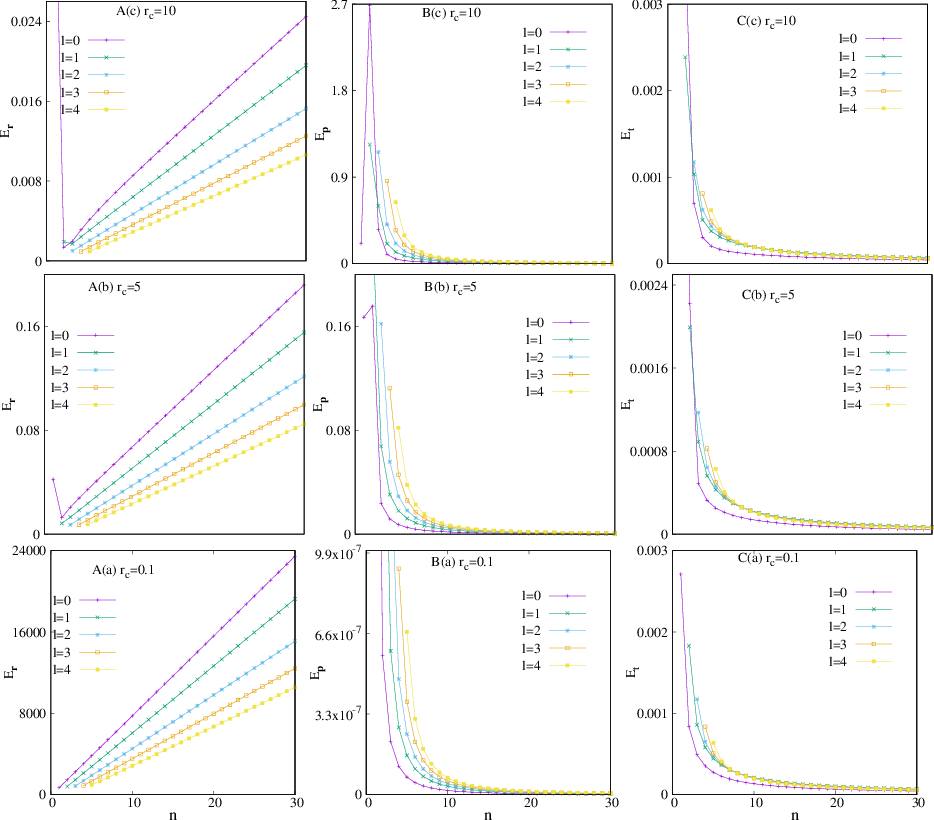}
\end{minipage}%
\caption{Plot of $E_{\rvec}$ (A), $E_{\pvec}$ (B) and $E_{t}$ (C) versus $n$ for $s,p,d,f,g$ 
orbitals at three particular $r_{c}$ values of CHA, namely, $0.1,5,10$ in panels (a)-(c).}
\label{oni_n_cha}
\end{figure}

$E$ exerts opposite effect to that of $R, S$. In case of a particular $n,l,m$ state, $E_{\rvec}$ diminishes and $E_{\pvec}$ 
escalates with enhancement in $r_{c}$. Table~\ref{oni_table_cha} presents $E_{\rvec},E_{\pvec},E_{t}$ for $1s, 2s$ at eight $r_{c}$, of CHA. 
As $r_c$ leads to zero, $E_{\rvec}$ follows the ordering $E_{\rvec}(5g)>E_{\rvec}(4f)>E_{\rvec}(3d)>E_{\rvec}(2p)$ which converses to 
$E_{\rvec}(2p)>E_{\rvec}(3d)>E_{\rvec}(4f)>E_{\rvec}(5g)$ in $r_c \rightarrow \infty$ limit. Contrariwise, $E_{\pvec}$ imprints exactly opposite 
feature from its $r$-space counterpart at both small and large $r_c$ regions. Now we can discern the reliance of $E$ on $n$. At a certain $l$ and 
$r_{c}=0.1$, $E_{\rvec}$, $E_{\pvec}, E_{t}$ assert completely opposite tendency to what is seen in $R,S$. The former advances and latter 
two decline with rise in $n$. This is demonstrated in bottom panels A(a), B(a), C(a) of Fig.~\ref{oni_n_cha} correspondingly. This again 
appreciates the fact that, effect of confinement prevails with advancement in $n$. It is appropriate to mention that, at $r_{c} \rightarrow \infty$, 
exactly inverse nature for $E_{\rvec}$ and $E_{\pvec}$ is attained. Here again, first two columns of middle and top rows of Fig.~\ref{oni_n_cha} 
suggest the appearance of maxima and minima in the graphs in $r,p$ spaces respectively. This supports that, with relaxation in confinement, CHA
proceeds towards FHA. These extrema will move forward towards right with rise in $r_{c}$. The effect of $l$ on $E$ is completely different to 
$R_{\rvec}^{\alpha},S_{\rvec}$. At a constant $n$, $E_{\rvec}$ abates and $E_{\pvec}$ escalates with progress in $l$. This has been discussed 
at length \cite{mukherjee18e} in the context of $n=10$ states. 

\begingroup           
\begin{table}
\small
\caption{$E_{\rvec}, E_{\pvec}$ for $1s$ and $2s$ states in CHA at some chosen $r_c$. See text for details.}
\centering
\begin{tabular}{>{\small}l>{\small}l>{\small}l|>{\small}l>{\small}l>{\small}l}
\hline
\multicolumn{3}{c}{$1s$}    \vline &      \multicolumn{3}{c}{$2s$}    \\
\cline{1-3} \cline{4-6}
$r_c$  &    $E_{\rvec}$           & $E_{\pvec}$         & $r_c$    &  $E_{\rvec}$    &     $E_{\pvec}$   \\ 
\hline 
0.1     &  685.24426269   &    0.00000395   &   0.1   &   1467.68253819     &  0.0000005   \\
0.2     &   87.40227398   &    0.00003142   &   0.2   &    185.27985826     &  0.0000045   \\
0.5     &    5.97242136   &    0.00047889   &   0.5   &     12.20852686     &  0.0000715   \\
1.0     &    0.84791755   &    0.00364537   &   1.0   &      1.59792065     &  0.0005811   \\
5.0     &    0.04217592   &    0.16706123   &   5.0   &      0.01264653     &  0.1756848   \\
7.5     &    0.03985512   &    0.20591605   &   7.5   &      0.00303307     &  1.0029980         \\
10.0    &    0.03978990   &    0.20886414   &   10.0  &      0.00135661     &  2.6889282         \\
40.0    &    0.03978873   &    0.20897494   &   40.0  &      0.00077712     &  7.6497493         \\ 
\hline
\end{tabular}
\label{oni_table_cha}
\end{table}
\endgroup

Four different complexity measures, given in Eq.~(\ref{complexity}) are probed with change of $r_c$, choosing two different and commonly 
used values of $b$ ($1,\frac{2}{3}$). Note that, for $b=1$, $C_{ES}^{(2)}$ modifies to $C_{LMC}$; analogously $C_{IS}^{(1)}$ coincides with 
$C_{IS}$ at $b=\frac{2}{3}$. In order to facilitate the discussion, a few words may be committed to the notation involved. A uniform symbol 
$C_{order_{s}, disorder_{s}}^b$ is used; where the two subscripts refer to two order ($E, I$) and disorder ($S, R$) parameters. 
Another subscript $s$ is used to signify the space; \emph{viz.}, $r, p$ or $t$ (total). Two scaling parameters $b=\frac{2}{3}, 1$ 
are illustrated with superscripts (1), (2). In this part we will discuss the outcome obtained from a recent complexity analysis 
undertaken by us \cite{majumdar17}. 

In CHA, it can be established that, out of these eight measures, $C_{ES}^{(2)}, E_{ER}^{(2)}$ and
$C_{IS}^{(1)}, E_{IR}^{(1)}$ can detect the presence of node in a given $n,l,m$ state. Because, in all these four measures,
there appear extrema (both in $r,p$ spaces) in nodal states. However, in $C_{ES}^{(1)}, E_{ER}^{(1)}$ and $C_{IS}^{(2)}, 
E_{IR}^{(2)}$ no such extrema is observed. Thus, it can be said that former two pairs explain CHA better in comparison to the latter
two pairs. Hence, depending upon the nature of complexity measures, it is required to determine a proper value of $b$.   

\subsubsection{Relative information for Hydrogen atom}
The radial wave function in $r$ space simplifies to commonly used form, as given below, 
\begin{equation} \label{irh}
\psi_{n,l}(r)= \frac{2}{n^2}\left[\frac{(n-l-1)!}{(n+l)!}\right]^{\frac{1}{2}}\left[\frac{2Z}{n}r\right]^{l} 
e^{-\frac{Z}{n}r} \ L_{(n-l-1)}^{(2l+1)} \left(\frac{2Z}{n}r\right).  
\end{equation}
$L_{(n-l-1)}^{(2l+1)}(s)$ is associated Laguerre polynomial. Putting $\xi=\frac{2Zr}{n}$ in Eq.~(\ref{ir}), one gets,
\begin{equation} \label{irh1} 
IR_{\rvec} = 4 \left(\frac{n}{2Z}\right) \int_{0}^{\infty} \xi^{2}R_{n,l}^{2}(\xi)\left[\frac{R_{n,l}^{\prime}(\xi)}
{R_{n,l}(\xi)}-\frac{R_{n_{1},l}^{\prime}(\xi)}{R_{n_{1},l}(\xi)}\right]^{2} \mathrm{d} \xi .
\end{equation}
Where primes denote 1st-order derivatives with respect to $\xi$. Then one can write, 
\begin{equation}
\frac{R_{n,l}^{\prime}(\xi)}{R_{n,l}(\xi)}=\frac{l}{\xi}-\frac{1}{2}-\frac{L_{(n-l-2)}^{(2l+2)}(\xi)}{L_{(n-l-1)}^{(2l+1)}(\xi)};  
\ \ \ \ \ \ \ \ \ \ \frac{R_{n_{1},l}^{\prime}(\xi)}{R_{n_{1},l}(\xi)}=\frac{l}{\xi}-\frac{1}{2}-
\frac{L_{(n_{1}-l-2)}^{(2l+2)}(\xi)} {L_{(n_{1}-l-1)}^{(2l+1)}(\xi)},
\end{equation}
since $R_{n_{1},l}(\xi)$ represents a circular state. So, $L_{(n_{1}-l-1)}^{(2l+1)}(\xi)$ provides a 
constant term, and hence $L_{(n_{1}-l-2)}^{(2l+2)}(\xi)=0$. This facilitate the second ratio as, 
$\frac{R_{n_{1},l}^{\prime}(\xi)}{R_{n_{1},l}(\xi)}=\frac{l}{\xi}-\frac{1}{2}$. Using above criteria and 
orthonormality condition of Associated Laguerre polynomial, Eq.~(\ref{irh1}) gives,
\begin{equation} 
IR_{\rvec} = \left(\frac{8}{Zn^3}\right)\frac{(n-l-1)!}{(n+l)!} \int_{0}^{\infty} \xi^{2l+2} e^{-\xi} 
\left[L_{(n-l-2)}^{(2l+2)}(\xi)\right]^{2} \mathrm{d} \xi,
\end{equation}    
which, after some algebraic manipulation, gives the simplified form of $IR_{\rvec}$ as below, 
\begin{equation} \label{irh2}
IR_{\rvec}=\frac{8 (n-l-1)}{Zn^3}  \ \ \ \ \ \ \ \  (\mathrm{when} \ \ n>l,  \ n-l \geq 2).
\end{equation}

Equation~(\ref{irh2}) clearly suggests that, $IR_{\rvec}$ reduces with rise of $n,~l$ and $Z$. Thus with progress in $n,~l$, spatial 
separation between distribution of reference circular and respective $l$ state decreases. In other words, fluctuation of a 
particular state with respect to node-less reference state reduces with increment of nodes. It may be recalled that the behavioral
pattern of $IR_{\rvec}$ with $n$ is similar to that of $I_{\rvec}$ \cite{romera05}--both diminish as $n$
advances. However, $I_{\rvec}$ is independent of $l$, whereas, $IR_{\rvec}$ seems to reduce with growth of $l$, for a fixed $n$. 
Table~\ref{tirh} presents some representative $IR_{\rvec}$ results for $n_ss, n_pp, n_dd, n_ff (n_s \geq 2, n_p \geq 3, n_d \geq 4, n_f \geq 5$)
orbitals for H atom, considering the respective circular states as reference.

\begin{figure}                         
\begin{minipage}[c]{1.0\textwidth}\centering
\includegraphics[scale=0.90]{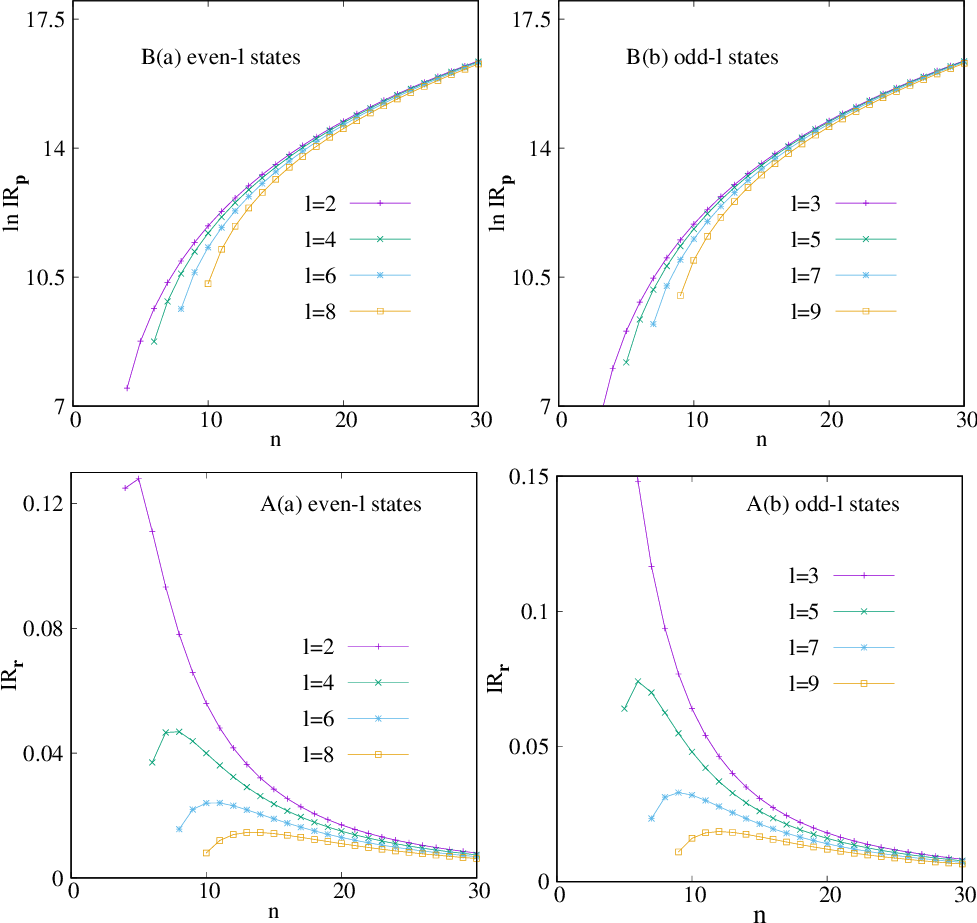}
\end{minipage}%
\caption{$IR_{\rvec}$ (A), $IR_{\pvec}$ (B) changes with $n$, for (a) even- (b) odd-$l$ states of H atom \cite{mukherjee18b}.}
\label{firh}
\end{figure}

In case of \emph{even}-$l$ states maximum in $IR_{\rvec}$ appears at $n=\frac{(3l+4)}{2}$, the resembling value being $\frac{32}{(3l+4)^3}(l+2)$. 
For \emph{odd}-$l$, the same occurs for $n=\frac{3}{2}(l+1)$, with a respective value of $\frac{32}{27}\frac{1}{(l+1)^2}$. The panels A(a)-A(b) 
of Fig.~\ref{firh} illustrate variations of $IR_{\rvec}$ with changes in $n$ for \emph{even} ($l=2,~4,~6,~8$) and \emph{odd} ($l=3,~5,~7,~9$) 
$l$ states respectively. Each curve passes through a maximum, which tends to shift towards right as $l$ attains higher values. One also notices 
that when $n\gg l$ then, we achieve $IR_{\rvec} \approx -\frac{16}{Z^3}\mathcal{E}_{n}$.

\begingroup           
\begin{table}
\small
\caption{Some specimen $IR_{\rvec}$ results for $n_ss, n_pp, n_dd, n_ff \ (n_s \geq 2, n_p \geq 3, n_d \geq 4, n_f \geq 5)$
orbitals of H atom, considering the corresponding circular states as reference \cite{mukherjee18b}. For $2s, 3s, 4s, 5s$ 
states, data given in parentheses represent literature results \cite{yamano18}.}
\centering
\begin{tabular}{c>{\small}l|c>{\small}l|c>{\small}l|c>{\small}l}
\hline 
Orbital   &  $IR_{\rvec}$                &  Orbital  &  $IR_{\rvec}$      &
Orbital   &  $IR_{\rvec}$                &  Orbital  &  $IR_{\rvec}$      \\  \hline
$2s$    & 1                (1)           &  $3p$   & $\frac{8}{27}$ & $4d$ &  $\frac{1}{8}$        &  $5f$ & $\frac{8}{125}$ \\
$3s$    & $\frac{16}{27}$  (0.5925)      &  $4p$   & $\frac{1}{4}$        & $5d$ &  $\frac{16}{125}$     &  $6f$ & $\frac{2}{27}$ \\
$4s$    & $\frac{3}{8}$    (0.375)       &  $5p$   & $\frac{24}{125}$    & $6d$ &  $\frac{1}{9}$     &  $7f$ & $\frac{24}{343}$ \\
$5s$    & $\frac{32}{125}$ (0.256)       &  $6p$   & $\frac{4}{27}$ & $7d$ &  $\frac{32}{343}$ &  $8f$ & $\frac{1}{6}$ \\
\hline
\end{tabular}
\label{tirh}
\end{table}
\endgroup

Next, we move on to $IR_{\pvec}$. The analytical expression \cite{sanudo08} for wave function reads,  
\begin{equation}
\psi_{n,l}(p)=n^{2}\left[\frac{2}{\pi}\frac{(n-l-1)!}{(n+l)!}\right]^\frac{1}{2} 2^{(2l+2)} \ l! \ 
\frac{n^l}{ \{[\frac{np}{Z}]^2+1 \}^{l+2}} \left(\frac{p}{Z}\right)^l
C_{n-l-1}^{l+1} \left(\frac{[\frac{np}{Z}]^2-1}{[\frac{np}{Z}]^2+1}\right), 
\end{equation}
where $C_{\zeta}^{\eta}(X)$ signifies the Gegenbauer polynomial. Let us consider $t=\frac{np}{Z},$ and subsequently
$q=\frac{t^2-1}{t^2+1}$. These two substitutions transform Eq.~(\ref{ir}) into the form,
\begin{equation}
IR_{\pvec} = Zl!^{2}n^{3}\left[\frac{2^{4l+7}}{\pi}\frac{(n-l-1)!}{(n+l)!}\right]
\int_{-1}^{1}(1+q)^{\frac{3}{2}}(1-q)^{\frac{5}{2}}[f_{n,l}(q)]^{2}
\left[\frac{d}{dq}\left(\frac{f_{n,l}^{\prime}(q)}{f_{n,l}(q)}-
\frac{f_{n_{1},l}^{\prime}(q)}{f_{n_{1},l}(q)}\right)\right]^{2}\mathrm{d}q,  
\end{equation}
where $f_{k,l}^{\prime}(q)=\frac{\mathrm{d}}{\mathrm{d}q}[f_{k,l}(q)]$,  
$f_{(k,l)}(q)=(1+q)^{\frac{l}{2}}(1-q)^{\frac{l}{2}+2}C_{k-l-1}^{l+1}(q).$ Also we note that,
\begin{equation}
\begin{aligned}
\frac{f_{n,l}^{\prime}(q)}{f_{n,l}(q)} & =\frac{l}{2(1+q)}-\frac{\frac{l}{2}+2}{1-q}+
\frac{\left(C_{n-l-1}^{l+1}(q)\right)^{\prime}}{C_{n-l-1}^{l+1}(q)} \\
\frac{f_{n_{1},l}^{\prime}(q)}{f_{n_{1},l}(q)} & =\frac{l}{2(1+q)}-
\frac{\frac{l}{2}+2}{1-q}+\frac{\left(C_{n_{1}-l-1}^{l+1}(q)\right)^{\prime}}{C_{n_{1}-l-1}^{l+1}(q)}. 
\end{aligned}
\end{equation}
Here too, as in case of $L_n^{\alpha}(x)$ for $IR_{\rvec}$, $C_{n_{1}-l-1}^{l+1}(q)$, being a part of circular state, is a 
constant and hence $(C_{n_{1}-l-1}^{l+1}(q))^{\prime}=0$. Therefore, one may write, 
\begin{equation}
\frac{f_{n_{1},l}^{\prime}(q)}{f_{n_{1},l}(q)}=\frac{l}{2(1+q)}-\frac{\frac{l}{2}+2}{1-q}.
\end{equation}
After going through some algebra, one gets the following expression, 
\begin{equation}
IR_{\pvec}=Z \ l!^{2} \ n^{3}\left[\frac{2^{4l+7}}{\pi}\ \frac{(n-l-1)!}{(n+l)!}\right](l+1)^{2}\left(I_{1}+I_{2}\right), 
\end{equation}
where the two integrations are defined as, 
\begin{equation}
I_{1}  =\int_{-1}^{1}(1-q^{2})^{l+\frac{3}{2}} \ \left[C_{n-l-2}^{l+2}(q)\right]^{2} \mathrm{d}q,  \ \ \ \ \ \ 
I_{2}  =\int_{-1}^{1}q(1-q^{2})^{l+\frac{3}{2}}\ \left[C_{n-l-2}^{l+2}(q)\right]^{2} \mathrm{d}q.
\end{equation}  
Using the fact that, $I_{2}=0$ as the integrand is an odd function of $q$, finally we obtain,  
\begin{equation}
\begin{aligned}
IR_{\pvec} & =Z \ l!^{2} \ n^{3}\left[\frac{2^{4l+7}}{\pi}\frac{(n-l-1)!}{(n+l)!}\right](l+1)^{2}I_{1} \\
           & = 2^{4} \ Z \ n^{2}[n^{2}-(l+1)^{2}] \ \ \ (\mathrm{when} \ n>l, \ n-l \geq 2).
\end{aligned}
\end{equation}
This equation suggests that, at a particular $l$, $IR_{\pvec}$ enhances with $n$. On the contrary, at a certain $n$, like $IR_{\rvec}$, 
it decreases with $l$. In panels B(a), B(b) of Fig.~\ref{firh}, $\mathrm{ln} \ (IR_{\pvec})$ is plotted against $n$ for the same set of even and 
odd-$l$ states (discussed in $IR_{\rvec}$ part) respectively. These two graphs conclude that, there is neither a maximum nor a minimum in 
$IR_{\pvec}$. In this case, when $n\gg l$, $IR_{\pvec} \approx \frac{4Z^{5}}{\mathcal{E}_{n}^{2}}$. 

$IR_{\rvec}$ measures the change of fluctuation from ground (fixed $l$) to excited states. Thus, in a nutshell it can be inferred that, result of 
$IR_{\rvec}$ in H atom provides more detailed information than $I$, as it incorporates the effect of $l$ in it \cite{romera05}. Not only that, it 
also reinforce the epilogue that, the diffused nature of an orbital with constant $l$ increases with increment of $n$.

\subsubsection{A virial-like theorem for CHA}
Recently a virial-like equation has been designed by the authors for confined quantum systems \cite{mukherjee19}, which takes the following form, 
\begin{equation} \label{eq:virial}
\begin{aligned}
\langle \hat{T}^{2} \rangle_{n}-\langle \hat{T} \rangle^{2}_{n} & = \langle \hat{V}^{2} \rangle_{n}-\langle \hat{V} 
\rangle^{2}_{n} \\
(\Delta \hat{T}_{n})^{2} & =   \langle \hat{V} \rangle_{n} \langle \hat{T} \rangle_{n}
-\langle \hat{T}\hat{V} \rangle_{n} = (\Delta \hat{V}_{n})^{2} = \langle \hat{T} \rangle_{n} \langle \hat{V} \rangle_{n}
-\langle \hat{V}\hat{T} \rangle_{n}.
\end{aligned}
\end{equation}
This relation concludes that, the magnitude of error associated with $\langle \hat{T} \rangle_{n}$ and $\langle \hat{V} \rangle_{n}$
are identical. Now, one can easily infer that, $\mathcal{E}_{n}$ is a sum of two average quantities but still provides 
exact result. It is because of the cancellation of errors between $\langle \hat{T} \rangle_{n}$ and $\langle \hat{V} \rangle_{n}$.
In a confinement condition, validity of Eq.~(\ref{eq:virial}) can be checked by deriving the expressions of 
$\langle \hat{T} \hat{V} \rangle_{n,\ell}$, $\langle \hat{V} \hat{T} \rangle_{n,\ell}$, $\langle \hat{V}^{2} \rangle_{n,\ell}$ and 
$\langle \hat{V} \rangle_{n,\ell}$ (other integrals remain unchanged). 

\begingroup           
\begin{table}
\small
\caption{$\mathcal{E}_{n,\ell}, \left(\Delta V_{n,\ell}\right)^{2}, \left(\Delta T_{n,\ell}\right)^{2}, \langle T \rangle_{n,\ell}
\langle V \rangle_{n,\ell}-\langle TV \rangle_{n,\ell}, \langle T \rangle_{n,\ell}\langle V \rangle_{n,\ell}-\langle VT 
\rangle_{n,\ell}$ of $1s,2s,2p$ states in CHA at four ($0.1,0.2,0.5,\infty$) $r_c$'s \cite{mukherjee19}. See text for detail.}
\centering
\begin{tabular}{>{\footnotesize}l|>{\footnotesize}l|>{\footnotesize}l>{\footnotesize}l>{\footnotesize}l>{\footnotesize}l}
\hline
State & Property           &  $r_c=0.1$ & $r_c=0.2$ & $r_c=0.5$ &  $r_c=\infty$  \\
\hline
      & $\mathcal{E}_{1,0}^{\P}$                                                        & 468.9930386  & 111.0698588 
& 14.74797003   & $-$0.499999 \\
      & $\left(\Delta V_{1,0}\right)^{2}$                                               & 308.8728899  & 80.38083598 
& 14.53962018     & 0.99999999 \\ 
$1s$    & $\left(\Delta T_{1,0}\right)^{2}$                                               & 308.8728899  & 80.38083598 
& 14.53962018      & 0.99999999 \\
      & $\langle T \rangle_{1,0}\langle V \rangle_{1,0}-\langle TV \rangle_{1,0}$ & 308.8728899  & 80.38083598 
& 14.53962018     & 0.99999999 \\
      & $\langle T \rangle_{1,0}\langle V \rangle_{1,0}-\langle VT \rangle_{1,0}$ & 308.8728899  & 80.38083598 
& 14.53962018     & 0.99999999 \\
\hline
      & $\mathcal{E}_{n,l}^{\dag}$                                                      & 1942.720354 & 477.8516723 
& 72.67203919   & $-$0.124999 \\
      & $\left(\Delta V_{2,0}\right)^{2}$                                               & 925.8428960 &  236.7351455 
& 40.51345969   & 0.18749999 \\ 
$2s$    & $\left(\Delta T_{2,0}\right)^{2}$                                               & 925.8428960 &  236.7351455 
& 40.51345969   & 0.18749999 \\
      & $\langle T \rangle_{2,0}\langle V \rangle_{2,0}-\langle TV \rangle_{2,0}$ & 925.8428960 &  236.7351455 
& 40.51345969   & 0.18749999 \\
      & $\langle T \rangle_{2,0}\langle V \rangle_{2,0}-\langle VT \rangle_{2,0}$ & 925.8428960 &  236.7351455 
& 40.51345969  & 0.18749999 \\
\hline
      & $\mathcal{E}_{2,1}^{\ddag}$                                                    & 991.0075894 & 243.1093321 
& 36.65887588   & $-$0.124999 \\
      & $\left(\Delta V_{2,1}\right)^{2}$                                              & 47.98046148 & 12.14249373     
& 2.016203448  & 0.02083333 \\ 
$2p$    & $\left(\Delta T_{2,1}\right)^{2}$                                               & 47.98046148 &  12.14249373    
& 2.016203448  & 0.02083333 \\
      & $\langle T \rangle_{2,1}\langle V \rangle_{2,1}-\langle TV \rangle_{2,1}$ & 47.98046148 &  12.14249373    
& 2.016203448  & 0.02083333 \\
      & $\langle T \rangle_{2,1}\langle V \rangle_{2,1}-\langle VT \rangle_{2,1}$ & 47.98046148 &  12.14249373    
& 2.016203448  & 0.02083333 \\
\hline   
\end{tabular}
\begin{tabbing} 
$^{\P}$Literature results \cite{roy15ijqc} of $\mathcal{E}_{1,0}$ for $r_{c}=0.1,0.2,0.5, \infty$ are: 
~468.9930386595,~111.0698588367,\\~14.74797003035,~$-$0.5 respectively. \\
$^{\dag}$Literature results \cite{roy15ijqc} of $\mathcal{E}_{2,0}$ for $r_{c}=0.1,0.2,0.5, \infty$ are: 
~1942.720354554,~477.8516723922,\\~72.67203919047,~$-$0.125 respectively. \\
$^{\ddag}$ Literature results \cite{roy15ijqc} of $\mathcal{E}_{2,1}$ for $r_{c}=0.1,0.2,0.5,\infty$ are: 
991.0075894412,~243.1093166600,\\~36.65887588018,~$-$0.125 respectively.
\end{tabbing}
\label{vch}
\end{table}
\endgroup

We begin with the \emph{exact} wave function for CHA given in Eq.~(\ref{wfh}). The pertinent expectation values can be simplified as, 
\begin{equation}
\langle \hat{T}\hat{V} \rangle_{n,\ell}= \langle \hat{T} v(r) \rangle_{n,\ell} + \langle \hat{T}v_{c}(r) \rangle_{n,\ell}= 
\langle \hat{T} v(r) \rangle_{n,\ell}. 
\end{equation}
In this instance,  $\langle \hat{T}v_{c}(r) \rangle_{n,\ell}=0$, as the wave function disappears for $r \geq r_c$. Use of same 
argument, along with the fact that $\langle v_{c}(r) \hat{T}\rangle_{n,\ell}=0$, gives rise to, 
\begin{equation}
\begin{aligned}
\langle \hat{V}\hat{T} \rangle_{n,\ell} & = & \langle v(r) \hat{T} \rangle_{n,\ell} + \langle v_{c}(r) \hat{T}  
\rangle_{n,\ell}= \langle  v(r) \hat{T} \rangle_{n,\ell}.  
\end{aligned}
\end{equation}  
Now, since $\langle v(r)v_{c}(r) \rangle_{n,\ell} = \langle v_{c}(r)v(r) \rangle_{n,\ell} = \langle v_{c}(r)^{2} 
\rangle_{n,\ell} =0$, one may write, 
\begin{equation}
\begin{aligned}
\langle \hat{V}^{2} \rangle_{n,\ell} = \langle v(r)^{2} \rangle_{n,\ell} + \langle v(r)v_{c}(r) \rangle_{n,\ell} + 
\langle v_{c}(r)v(r) \rangle_{n,\ell} + \langle v_{c}(r)^{2} \rangle_{n,\ell} =  \langle v(r)^{2} \rangle_{n,\ell}. 
\end{aligned}
\end{equation}
Again, because  $\langle v_{c}(r) \rangle_{n,\ell}=0$, it follows that, 
\begin{equation}
\langle \hat{V} \rangle_{n,\ell} = \langle v(r) \rangle_{n,\ell} + \langle v_{c}(r) \rangle_{n,\ell} = \langle v(r) 
\rangle_{n,\ell}. 
\end{equation}

Thus, for a CHA also, Eq.~(\ref{eq:virial}) remains unchanged, i.e., 
\begin{equation}
\begin{aligned}
\langle \hat{T}^{2} \rangle_{n,\ell}-\langle \hat{T} \rangle^{2}_{n,\ell} & = \langle \hat{V}^{2} \rangle_{n,\ell}-\langle 
\hat{V} \rangle^{2}_{n,\ell} \\
\left(\Delta \hat{T}_{n,\ell}\right)^{2} = \left(\Delta \hat{V}_{n,\ell}\right)^{2} & = \langle \hat{T} \rangle_{n,\ell} 
\langle v(r) \rangle_{n,\ell}
-\langle v(r)\hat{T} \rangle_{n,\ell} = \langle \hat{T} \rangle_{n,\ell} \langle v(r) \rangle_{n,\ell}
-\langle \hat{T} v(r) \rangle_{n,\ell}.
\end{aligned}
\end{equation}
This equation suggests that, CHA fulfills the results given in Eq.~(\ref{eq:virial}); and $v_c$ has no impact on it. It has only 
introduced the boundary in a finite range. Table~\ref{vch} demonstrates sample values of $\mathcal{E}_{n,\ell}$, 
$(\Delta \hat{T}_{n,\ell})^{2}$, $(\Delta \hat{V}_{n,\ell})^{2}$, $\langle T \rangle_{n,\ell}\langle V 
\rangle_{n,\ell}-\langle TV \rangle_{n,\ell}$ and $\langle T \rangle_{n,\ell}\langle V \rangle_{n,\ell}-\langle VT 
\rangle_{n,\ell}$ for same low-lying ($1s,~2s,~2p$) states, in CHA at four particular $r_c$'s, like 
$0.1,~0.2,~0.5,~\infty$. For sake of completeness, accurate values of $\mathcal{E}_{n,\ell}$ are reproduced from 
\cite{roy15ijqc}. In both confining and free (last column) conditions, these results complement the conclusion of 
Eq.~(\ref{eq:virial}). In passing, it is interesting to note that both $(\Delta \hat{T}_{n,\ell})^{2}, 
(\Delta \hat{V}_{n,\ell})^{2}$ decrease with rise in $r_c$.

\subsection{Many-electron atom in spherical cage}
In this subsection, we briefly cover some fundamental aspects regarding the application of DFT to many-electron atoms, where the 
electrons remain under the influence of an external potential that imposes restriction of an impenetrable spherical cavity. DFT has been 
successfully implemented in explaining the electronic structure and properties of many-electron systems in their ground state. It is, however, noted 
in the derivation of Hohenberg-Kohn theorems that, DFT is usually termed as a ground-state theory as the practical approach to extract the information 
on excited-state energies and densities has remained a bottleneck. About a decade ago, an exchange potential based on 
the work-function approach \cite{sahni92} was proposed to perform both ground and excited state calculations in many-electron atoms.  
This non-variational method is simple, computationally efficient, and has shown considerable success for some singly, doubly, and triply excited 
states, low and moderately high states, valence and core excitations, as well as in explaining the auto-ionizing and satellite states for free 
systems \cite{roy97, roy97a, roy97b, roy98, roy99a, roy02b, roy04jpbhollow, roy05jpbhollow, roy07}. However, while the success of this method 
is well-established and well documented, it has never been tested so far, in the context of confinement studies, such as those undertaken here. 
Therefore, it will be a very much worthwhile investigation. Due to the existence of 
Coulomb singularity at the origin and its long-range nature, we employ the GPS procedure to achieve high accuracy. It allows nonuniform 
spatial grid discretization and considerably more accurate calculations of eigenvalues and wave functions can be obtained using only a modest number 
of grid points. Here we shall present some new complementary results for further evaluation of the methodology in regards to \emph{hard} 
confinement. 

\subsubsection{Spherically confined atoms}
In order to see the contribution of each part of the exchange-correlation (XC) functionals, in Table 8 we present the X-only energy of confined 
He atom in its ground state obtained by work-function approximation as a function of confining radius and they are contrasted with the HF results 
reported in \cite{montgomery13}. The second and third columns of this table stand for the X-only results calculated with our method and HF results
respectively. Also total energy considering the local and gradient corrected correlation functionals are reported in fourth and fifth columns. 
Next two columns quote energies reported in the literature obtained with wave function based-approaches; the multi-configuration 
parametrized optimized effective potential (JPOEP) method \cite{sarsa16} and variational energies calculated with Hylleraas wave function 
\cite{garza11}. From this table the appreciable conclusions which can be drawn are: i) results obtained 
from our proposed method indicate the correct behavior of total energy as a function of confinement radii when compared with corresponding 
available information for He, and ii) our X-only results are in good agreement with HF values. For all $r_c$'s the X-only energies match up to 
four decimal points with HF energies. Predictably, the local correlation functionals (here Wigner) perform worse than the gradient-corrected 
(LYP) one. Considering Hylleraas wave function to be the best variational wave function for two-electron systems, we may conclude that LYP functional 
works well for weak confinements but suffers in small $r_c$ region, which needs to be addressed in future. 

\begingroup           
\begin{table}
\small
\caption{Ground-state energy of confined He at some chosen $r_c$. See text for details.}
\centering
\begin{tabular}{>{\footnotesize}l|>{\footnotesize}l>{\footnotesize}l|>{\footnotesize}l>{\footnotesize}l>{\footnotesize}l>{\footnotesize}l}
\hline
$r_c$  &  $E_{X}$ & $E_{HF}$ & $E_{XC}${\scriptsize(Wigner)}    & $E_{XC}${\scriptsize(LYP)}   &  $E${\scriptsize(JPOEP)}  & 
$E${\scriptsize(Hylleraas)}   \\ 
\hline 
0.1     &  906.61645  &  -----     &  906.4496   & 907.455   &  ----    &   906.562423     \\
0.5     &  22.79096  &   22.79095  &  22.6909   &  22.9592  &   ----    &    22.741303     \\
0.8     &  4.65737  &    4.65739   &  4.5787   &   4.7289 &   ----      &     ----         \\
1.0     &  1.06121  &    1.06122   &  0.9915   &   1.0988 &   1.0158    &     1.015755     \\
2.0     & $-$2.56256  &   $-$2.56251  &  $-$2.6115   &  $-$2.5879  & $-$2.6040  &   $-$2.604038      \\
3.0     & $-$2.83103  &   $-$2.83102  &  $-$2.8746   &  $-$2.8688  & $-$2.8724   &  $-$2.872494       \\
5.0     & $-$2.86136  &   $-$2.86129  &  $-$2.9035   &  $-$2.9033  & $-$2.9033   &  $-$2.903410       \\
100.0   & $-$2.86164  &   $-$2.86164  &  $-$2.9037   &  $-$2.9053  & $-$2.9036   &  $-$2.906956      \\ 
\hline
\end{tabular}
\label{ene_table_che}
\end{table}
\endgroup

We have particular interest in assessing the accuracy and reliability to describe the excited states of confined many-electron atom, 
for which literature results are quite scanty. In Table 9, the interplay between correlations and confinement is analyzed for first 
excited ($1s2s \ ^{3}S$) state of He. These energies were calculated over a range of $r_c$ varying from 0.1 to 10.0 (in a.u.). Selected energies 
from this work, with Wigner and LYP correlation functionals, and those calculated with Hylleraas wave function from \cite{montgomery13} 
are recorded in second, third and fourth columns respectively. It is clear that both methods predict a marked increase in energy, 
as we reduce the confinement radius. An analysis of XC potential for this particular case shows that, both the correlation energy functionals
can provide reasonable good agreements with other sophisticated wave-function-based approaches (considering energies for Hylleraas to be the 
most authentic one).

\begingroup           
\begin{table}
\small
\caption{Single excitation (1s2s, $^{3}S$) energy in confined He at some chosen $r_c$.}
\centering
\begin{tabular}{>{\footnotesize}l|>{\footnotesize}l>{\footnotesize}l>{\footnotesize}l>{\footnotesize}l}
\hline
$r_c$  &    $E_{X}$ & $E_{XC}${\scriptsize(Wigner)}         & $E_{XC}${\scriptsize(LYP)}   &  $E${\scriptsize(Hylleraas)}     \\ 
\hline 
0.1        &  2370.7389 &  2370.5729 &  2372.1569 & 2370.7270     \\
0.5        &  78.83157  &   78.7330  &    79.1755 &   78.8208     \\
1.0        &  14.36896   &  14.3014  &    14.4954 &   14.3597     \\
2.0        &   0.56698  &    0.5227  &     0.5795 &    0.5602     \\
3.0        &  $-$1.36562  &   $-$1.4009  &    $-$1.3822 &   $-$1.3705      \\
5.0        &  $-$2.04515  &   $-$2.0740  &    $-$2.0732 &   $-$2.0480      \\
10.0       &  $-$2.17079  &   $-$2.1964  &    $-$2.1985 &   $-$2.17262     \\ 
\hline
\end{tabular}
\label{enex_table_che}
\end{table}
\endgroup

\section{Conclusion}
In this chapter, at first we have summarized the role of $I,S,E$ and Onicescu-Shannon complexity in understanding the quasi-degeneracy and
trapping of a particle within either of the wells in an SDW potential. Later, we employed the same information measures to
explore oscillatory nature of a particle in an ADW potential. A thorough analysis exposes that, in SDW potential, conventional uncertainty
measures are unable to explain the various rich facets (such as quasi-degeneracy, localization of particle) associated with it, in a satisfactory
manner. However, investigation of $S, E$ suggests that, there occurs an interplay between localization 
and delocalization. At first delocalization predominates with increase in $S_{x}$ and decrease in $E_{x}$, but after a characteristic $\beta$, 
this trend reverses--localization prevails. It is imperative to mention that, in this context, $S_{x}$ passes through a maximum and 
$E_{x}$ comes out from a minimum. In case of ADW potential, a pair of rules have been demonstrated to discern quasi-degeneracy and 
to-and-fro motion of a particle between deeper and shallower wells. It also suggests that (i) a particle in an $n$th state oscillates $n$ times 
before localizing in deeper well (ii) at certain sets of $\alpha, \beta, \gamma$, ADW behaves as two different potentials.

Next, the effect of moving the boundary from infinity to finite region has been delineated in CHA through numerous information quantities quite
decently uncovering many hitherto unreported interesting features. A systematic and elaborate survey leads to the fact that, at sufficiently low 
$r_{c}$, the patterns in $I,S,R,E$ get reversed to that encountered in FHA. This interprets that, the effect of confinement is more on higher 
quantum states. Apart from that, the feasibility and accessibility of a new virial-like theorem, recently proposed by us, was tested in this context.
As expected, this turns out to be quite successful and physically meaningful. Further, relative Fisher information for FHA are also
reported here. Lastly, we inspected the influence of similar boundary condition on ground and excited states of helium atom using
a work-function based recipe within DFT. This proposed new method offers correct physical behavior in confinement in both ground and excited states. 
The illustrative results presented here for the first time, demonstrate its promise, and may be pursued for other many-electron atoms, in future. Two
representative correlation energy functionals were put to test to account for the subtle effects of electron correlation. While they produce 
better energies, from exchange-only calculations, there is clearly a need for more accurate and reliable functionals, especially 
in stronger confinement region. On the other side, however, it is abundantly clear that, this single-determinantal approach offers very competitive 
excited-state energies compared to other wave function-based multi-configuration or basis-set-related sophisticated methods. In this case, only energy 
analysis has been provided. Information 
analysis for confined helium  as well as other atoms will be undertaken in due course. In future, it will be desirable to explore the effect of more 
realistic (penetrable) boundary on such systems. Further, investigation of relative information in confined atoms using various reference states would 
be highly useful.

\section{Acknowledgement}
The authors sincerely thank Prof. Nadya S. Columbus, President, NOVA Science Publishers, NY, USA, and the Editor Dr. Vijay Kumar, for the kind invitation to 
contribute a chapter in this topical area. AKR gratefully acknowledges financial support from SERB INDIA, New Delhi (sanction no. CRG/2019/000293, 
MTR/2019/000012) and CSIR INDIA (sanction no. 01/(3027)/21/EMR-II).  

\bibliographystyle{unsrt} 
\bibliography{ref}
\end{document}